\documentclass{getwriting}

\usepackage{enumerate}
\usepackage{fancyhdr}
\usepackage{bm}
\usepackage{algorithm}
\usepackage{graphicx}
\usepackage[noend]{algpseudocode}
\usepackage{mathrsfs}
\usepackage{amsmath}
\usepackage{icomma}
\usepackage{float}
\newtheorem{proposition}{Proposition}
\usepackage{bbm}
\newcommand*{\rom}[1]{\expandafter\@slowromancap\romannumeral #1@}
\newcommand{\mathleft}{\@fleqntrue\@mathmargin0pt}
\usepackage[style=authoryear-icomp,maxbibnames=9,maxcitenames=1,backend=biber]{biblatex}
\usepackage{caption}
\usepackage{wrapfig}

\usepackage{amsmath}
\usepackage{amssymb}

\def\BREM{{\sc BREM}}

\begin{document}

\title{Bayesian Reconstruction and Differential \\ Testing of Excised mRNA}



\author[1]{Marjan Hosseini$^{*,}$}
\author[1]{Devin McConnell}
\author[1]{Derek Aguiar\thanks{Corresponding authors \{marjan.hosseini,derek.aguiar\}@uconn.edu}$^{,}$}
\affil[1]{\footnotesize Department of Computer Science \& Engineering University of Connecticut, Storrs CT 06269, USA}

\maketitle              

\abstract{
Characterizing the differential excision of mRNA is critical for understanding the functional complexity of a cell or tissue, from normal developmental processes to disease pathogenesis.
Most transcript reconstruction methods infer \textit{full-length transcripts} from high-throughput sequencing data.
However, this is a challenging task due to incomplete annotations and the differential expression of transcripts across cell-types, tissues, and experimental conditions.
Several recent methods circumvent these difficulties by considering \textit{local splicing events}, but these methods lose transcript-level splicing information and may conflate transcripts.  
We develop the first probabilistic model that reconciles the transcript and local splicing perspectives. 
First, we formalize the sequence of mRNA excisions (SME) reconstruction problem, which aims to assemble variable-length sequences of mRNA excisions from RNA-sequencing data.
We then present a novel hierarchical \textbf{B}ayesian admixture model for the \textbf{r}econstruction of \textbf{e}xcised \textbf{m}RNA (\BREM{}). 
We demonstrate the compactness of our probabilistic model by computing a minimum node-cover of a graph associated with splicing complexity in polynomial time, develop posterior inference algorithms based on Gibbs sampling and local search of independent sets, and characterize differential SME usage using generalized linear models based on converged \BREM{} model parameters. 
\BREM{} interpolates between local splicing events and full-length transcripts and thus focuses only on SMEs that have high probability in the posterior.
We show that \BREM{} achieves higher recall, F1 score, and accuracy for reconstruction tasks and improved accuracy and sensitivity in  differential splicing when compared with four state-of-the-art transcript and local splicing methods on simulated data.
Lastly, we evaluate \BREM{} on both bulk and scRNA sequencing data based on transcript reconstruction, novelty of transcripts produced, model sensitivity to hyperparameters, and a functional analysis of differentially expressed SMEs, demonstrating that \BREM{} captures relevant biological signal. 
The source code for \BREM{} is freely available at \href{https://github.com/bayesomicslab/BREM}{https://github.com/bayesomicslab/BREM}.


 \begin{refsection}

\section{Introduction}\label{aba:sec1}

Alternative splicing (AS) is characterized by the excision of pre-mRNA segments (typically intronic RNA) by the RNA-protein spliceosome complex and enables a single gene to produce multiple distinct and functionally diverse protein isoforms~\parencite{wilkinson2020rna}.
Alternative splicing is both prevalent, affecting an estimated 95\% of human protein-coding genes~\parencite{pan2008deep}, and integral for human adaptation~\parencite{barbosa2012evolutionary,keren2010alternative,merkin2012evolutionary}, gene regulation and tissue identity~\parencite{boudreault2016global,kornblihtt2013alternative,barbosa2012evolutionary,baralle2017alternative}, and disease etiology and drug resistance~\parencite{tazi2009alternative,lee2016therapeutic,yang2019aberrant}. 
Given the importance of AS in developmental biology and disease etiology, considerable effort has been devoted to computationally infer both the structure and expression of alternatively spliced transcripts across differing cellular contexts. 
High-throughput single cell and bulk RNA sequencing (scRNA-seq and RNA-seq respectively) provide experimental platforms for discovering and quantifying alternative splicing from short-read data.
After sequencing, reads are typically mapped to a reference genome with a splice-aware aligner that accounts for intronic gaps in the read alignment~\parencite{dobin2013star}.
Reads that map to a region for which intronic RNA was removed (splice junctions) are informative of the latent transcript diversity of the sample and can be assembled into putative transcripts \textit{de novo} or with reference transcriptome annotations.
Quantification is then determined as a function of the number of reads mapped to a specific transcript.

However, the computational characterization of AS is challenging due to biological variability and technological limitations.
First, both the structure and frequency of spliced transcripts (hereafter, \textit{transcripts} for brevity), differ by population, sex, tissue, and cell type \parencite{blekhman2010sex,ongen2015alternative,lappalainen2013transcriptome,park2018expanding,gtex2020gtex}.
Second, the structure of transcripts is often unknown or incomplete for many cell types, cell states, or non-model organisms~\parencite{morillon2019bridging}.
Third, transcripts have significant overlap of both retained and excised sequence making it difficult to distinguish the transcript of origin from short-read sequencing.
Lastly, the short read-lengths of high-throughput sequencing technologies limit the number of splice junctions observed in any single observation. 
Long-read sequencing technologies yield observations with many more splice junctions but suffer from higher costs, larger error rates, and lower throughput~\parencite{mantere2019long}.

Despite these challenges, a significant number of isoform reconstruction and quantification methods have been developed with different modelling assumptions and reconstruction goals~\parencite{aguiar2018bayesian,vaquero2016new,li2018annotation,trapnell2013differential}.
Among these annotation-based methods, the majority reconstruct \textit{full-length} transcripts defined by their \textit{composite exons}.
The Bayesian isoform discovery and individual specific quantification (BIISQ) method models transcript reconstruction with a nonparametric Bayesian hierarchical model, where samples are mixtures of transcripts sampled from a population transcript distribution~\parencite{aguiar2018bayesian}.
While BIISQ was shown to have high accuracy on low abundance isoforms, it requires both the genes and the composite exon coordinates, and is unable to construct isoform transcripts that deviate from this reference annotation.
Cufflinks and StringTie are two methods that construct full-length transcripts and can operate both with or without transcript annotations. 
Cufflinks reconstructs transcripts as minimum paths in an associated graph, where the aligned reads are vertices, and edges denote the compatibility of isoforms~\parencite{trapnell2010transcript}.
StringTie models transcript reconstruction using maximum network flow on a splice graph, where paths and read coverage inform isoform composition and quantification respectively~\parencite{pertea2015stringtie}.
Both are well-established state-of-the-art methods, but consider samples individually during the initial reconstruction.
For many genes, this reconstruction problem is underdetermined, uncertainty in 5' or 3' splice sites make it difficult to identify constituent exons, and variability of read depths due to technical artifacts or biological biases obfuscates reconstruction and quantification~\parencite{mcintyre2011rna}. 
In fact, full-length transcripts can be difficult to reconstruct and quantify even when transcriptome annotations are known~\parencite{vaquero2016new}. 

\begin{figure*}[h!]
\centering
\includegraphics[width=1\textwidth]{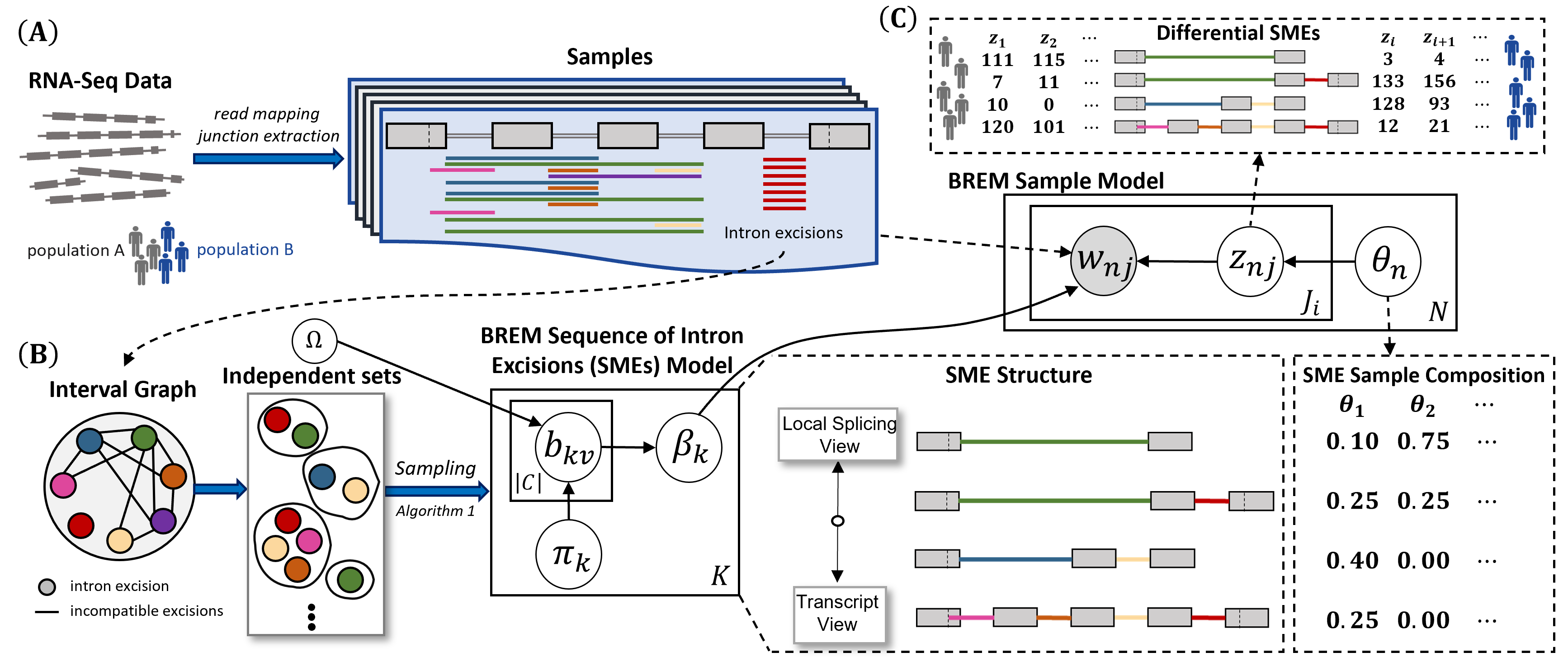}
\caption{\textbf{\BREM{} overview.} 
(A) Junction reads are extracted from short-read RNA-seq data and (B) used to construct an interval graph where nodes are mRNA excisions and edges connect two mRNA excisions if they overlap. 
\BREM{} is an admixture model where the sequences of mRNA excisions (SMEs) is informed by the interval graph.
(C) Posterior inference algorithms yield SMEs and counts of reads mapped to SMEs, which are used to compute differential SME usage.
Local splicing methods would conflate the first two transcripts while full-length methods maybe struggle differentiating the last two transcripts due to their large sequence overlap.
In both cases, these issues may affect differential splicing estimates.
}
\label{fig:overview}
\end{figure*}

Recently, several methods have been proposed that focus on the RNA that is excised from pre-mRNA transcripts and relax the requirement of reconstructing full transcripts.
The local splicing and hierarchical model rMATS detects differential usage of exons through the comparison of exon-inclusions in junction reads among five different alternative splicing events~\parencite{shen2014rmats}. 
Interestingly, LeafCutter focuses on the mRNA that is excised rather than the constituent exons of a transcript to identify local splicing events. 
First, LeafCutter computes local splicing events from RNA-Seq data then constructs a graph $G_L = (V_L,E_L)$ where vertices, $V_L$, are excisions and edges, $E_L$, connect excisions that share a donor or acceptor splice site~\parencite{li2018annotation}.
Subsequently, differential splicing of excised sequences in the connected components of $G_L$ is computed using a Dirichlet-Multinomial generalized linear model on read counts.
LeafCutter does not suffer the same disadvantages of methods that use exonic sequences or attempt to reconstruct full-length transcripts, though at the expense of the inability to identify certain splicing events like alternative transcription start sites. 
These methods are ideal for emerging technologies like scRNA-seq and tissue or disease specific splicing, since these applications suffer from low coverages and incomplete annotations; instead, they use sequences that overlap mRNA excisions (or \textit{junctions}), which are much more easily inferred by short read sequences at lower coverages and without transcript or exon annotations~\parencite{li2018annotation,gtex2020gtex}. 
While these \textit{local} splicing methods do not suffer from the same disadvantages as the \textit{transcript-based} methods, they are limited to singular splicing events (or small neighborhoods around an event)~\parencite{li2018annotation,shen2014rmats,vaquero2016new}.
As a result, these methods may conflate multiple transcripts that share splice events, making (a) quantification and downstream haplotype analysis difficult, and (b) differential expression subject to ambiguity of transcript-level contributions~(Fig.~\ref{fig:overview}). 

Here, we propose a hierarchical \textbf{B}ayesian admixture model for the \textbf{r}econstruction of \textbf{e}xcised \textbf{m}RNA (\BREM{}), a novel approach to isoform reconstruction and differential expression.
Unlike Cufflinks and StringTie, \BREM{} considers all samples jointly in a formal probabilistic model; further, \BREM{} does not require exon or transcript-level annotations and enables \textit{local}-to-\textit{full-length} transcript reconstruction~(Fig.~\ref{fig:overview}).
First, we define the \textit{sequence of mRNA excisions (SME) reconstruction problem}, focusing on the assembly of excised mRNA into sequences of mRNA excisions (hereafter, simply \textit{excisions}) from RNA-seq data.
Then, we develop a novel hierarchical Bayesian admixture model to solve the SME reconstruction problem and a differential splicing workflow based on model parameter estimates in a generalized linear model; admixture refers to samples being modelled as collections of sequence reads that are themselves sampled from global mixture components (SMEs). 
We demonstrate the theoretical compactness of \BREM{} and develop Gibbs Sampling and local search-based inference algorithms that model the discovery of new SMEs as computing independent sets in an interval graph.
We demonstrate increased precision, recall, and F1 score for transcript reconstruction on simulated data using both well-established and novel measures and highly accurate and sensitive differential SME identification. 
Lastly, we evaluate \BREM{} on both bulk and scRNA sequencing data based on transcript reconstruction, novelty of transcripts produced, sensitivity to hyperparameters, and a functional analysis of differentially expressed SMEs, demonstrating that \BREM{} captures relevant biological signal.

\section{Methods}\label{sec:met}

We are given RNA sequencing data $\mathcal{D}$ that has been aligned to a reference genome. 
Let the samples be indexed by $i$ for $i \in \{1,\dots,N\}$ and the total number of junction reads per sample be $J_i$.
Though a non-trivial problem, we assume aligned sequence reads can be assigned to a specific gene for ease of exposition. 
The \textbf{isoform reconstruction problem} aims to reconstruct, for each sample $i$, the full-length isoform transcripts as defined by their component exons.
Reads that overlap exon junctions (i.e. \textit{junction reads}) are highly informative for transcript reconstruction.

In contrast, the \textbf{splice event reconstruction problem} aims to identify singular splicing events that exist in any transcript expressed in $\mathcal{D}$. 
Since there need not be assembly of transcripts, this problem is generally a simpler computational task than the \textit{isoform reconstruction problem}, but can still yield biologically interesting insights, e.g., differential usage of particular splice sites.

Here, we introduce the \textbf{SME reconstruction problem}: given aligned RNA-seq data $\mathcal{D}$, reconstruct sequences of co-occurring excised mRNA.
This problem interpolates between the isoform reconstruction problem and the splice event reconstruction problem as special cases: i.e., when the sequences are defined as full-length transcripts or singular splice sites.
Since methods that focus on local splicing events cannot compute reliable transcript abundances, we only consider the problem of differential expression between two groups. 
Whenever the context is clear, we will refer to both differential usage of local splicing events, transcripts, and SMEs simply as \textit{differential expression}.

\subsection{\BREM{}:  \textbf{B}ayesian \textbf{R}econstruction of \textbf{E}xcised \textbf{m}RNA}
\label{sec:modeldescription}

\BREM{} solves the SME reconstruction problem by representing samples as mixtures of SMEs sampled from a global distribution that is learned across samples (Fig.~\ref{fig:overview}). 
SMEs are sequences of mRNA excisions, typically, but not limited to introns, that can be assigned to the same transcript (i.e., they do not overlap). 
Briefly, \BREM{} models samples as mixtures of SMEs, which are themselves mixtures of junction reads.
\BREM{} learns the structure of SMEs, a global distribution over SMEs, a mapping between junctions reads and SMEs, and a sample-specific distribution of SMEs. 
A separate model is built for each gene, which includes $i=1 \dots N$ samples that are collections of reads overlapping excision junctions. 
Let the set of unique excisions within a gene be $V$, which is indexed by $v \in \{1, \dots, |V|\}$.
The $i^{th}$ sample consists of $1, \dots, J_i$ junction reads. 
The goals are to reconstruct the latent SMEs and assign sample reads to SMEs for subsequent differential testing of SMEs. 
\BREM{} consists of two major components: (a) combinatorial model for SME structure and (b) a probabilistic model for SME admixture.

\subsubsection{Combinatorial model for SME structure.}
\label{combstruct}
We represent excisions as intervals on the genome, defined as tuples: \textit{(start position, terminal position)} and SMEs as sequences of excisions.
The goal is to arrange excisions into $K$ SMEs, such that no SME contains two overlapping excisions.
To enforce this criteria, we compute a graph $G = (V, E)$, with $v \in V$ for unique excision $v$ and $(v_1, v_2) \in E$ if $v_1 = (s_1, t_1)$ and $v_2 = (s_2, t_2)$ intersect, i.e. $\min(t_1, t_2) - \max(s_1, s_2) > 0$.
Note that independent sets in this graph correspond to valid SMEs for which no pair of excisions overlap.

In our probabilistic model, we enforce that two excisions should not be expressed in the same SME if they are connected in $G$ using Bernoulli random variables.
For instance, if $v_1$ and $v_2$ are excisions connected in $G$, then we can enforce $v_1 \oplus v_2$ where $\oplus$ is exclusive OR.
We create a Bernoulli random variable $b_{kv_1}$, where $b_{kv_1}=1$ if $v_1$ is in the $k^{th}$ SME and $0$ otherwise. 
Similarly $(1-b_{kv_1})$ is $1$ if $v_2$ is present and $0$ otherwise.
Unfortunately, this strategy does not scale well, as the number of random variables in the probabilistic model would be proportional to the number of conflicts.
For example, consider the complete bipartite graph $K_{1,k}$ (or, the star $S_k$). 
This tree has a single internal node and $k$ leaves, which would generate $k$ Bernoulli random variables because there are $k$ conflicts. 
However, notice that since the internal node is connected to all leaves, if the excision represented by the internal node is selected to be in the SME, none of the leaves may be added and the model can be described with a single random variable.

A parsimonious representation of SMEs reduces the number of parameters in our model, making inference more efficient and mitigating issues associated with model non-identifiability. 
Consider the excision graph as defined earlier: $G=(V,E)$ where $v \in V$ for unique excision $v$ and $(v_1,v_2) \in E$ if $v_1$ and $v_2$ intersect.
Edges represent excisions that cannot be co-expressed in the same SME.
Let $C$ be the set of Bernoulli random variables required to encode all conflicts in $G(V,E)$. 

\begin{proposition}
\label{theo:mvc}
Choosing the minimum number of variables required to encode all conflicts between excisions i.e., computing $C$ such that $|C|$ is minimum, can be done in time $O(|E|)$. 
\end{proposition}


Since each edge in $E$ denotes two excisions that cannot coexist in the same cluster, we need at least one incident node of each edge to exist in $C$; this is a node cover of $G$. 
A minimum node cover has the smallest cardinality among all node covers and therefore a corresponding $C$ for which $|C|$ is the smallest.
Since $G$ is an interval graph, computation of a minimum node cover can be done in $O(|E|)$ time~\parencite{marathe1992generalized}.

\subsubsection{Probabilistic model for SME admixture.}

The probabilistic component of \BREM{} consists of a model for SME structure that is shared across all samples (Fig.~\ref{fig:overview}, B and Fig.~\ref{fig:graphicalmodel}, left) and a model for the SME composition of a specific sample (Fig.~\ref{fig:overview}, C and Fig.~\ref{fig:graphicalmodel}, right); complete model details can be found in \S \ref{supmoddets}.
The structure of an SME consists of the inclusion or exclusion of excisions. 
We place an explicit beta-Bernoulli prior on excisions to control the sparsity of SMEs.
\begin{align*}
    b_{kv} & \sim Bernoulli(\pi_k), \forall v\in C \\
      s.t. & \hspace{20pt} \bm{b_{k\cdot}} \in \Omega \\
    \pi_k & \sim Beta (r,s)
\end{align*}
for hyperparameters $r$ and $s$ and the space of valid SMEs $\Omega$, which is defined by all subsets of excisions in which there exists no two elements that conflict in $G$; thus, $\Omega$ is equivalent to the \textit{set of all (not necessarily maximal) independent sets in $G$}. 
In total, we only instantiate $|C|$ Bernoulli variables since we can encode all $v \in C$ using $b_{kv}$ and all $\hat{v} \notin C$ with variables of the form $(1-b_{kv})$ for some $v \in C$.
If a single $\hat{v} \notin C$ is adjacent to two or more $v \in C$, one adjacent $v$ is selected at random for the encoding.
Importantly, $r$ and $s$ can be adjusted to encourage shorter or longer SMEs by affecting the prior probability of excision inclusion (see \S \ref{trecon}).

We model the $k^{th}$ SME, $\beta_k$, as a degenerate Dirichlet distribution whose dimension is controlled by the beta-Bernoulli prior~\parencite{wang2009decoupling,aguiar2018bayesian}.
Intuitively, we discourage excisions to occupy the same SME based on the structure of the excision interval graph $G$. 
In SME $k$, we enforce these constraints through the $|V|$-dimensional $\bm{b_k} = (b_{k1}, \dots, b_{k|V|})$ vector, in which $b_{kv}$ selectively turns off or on dimension $v$. 
Note that there are only a total of $|C|$ unique $b_{kv}$ as some of these variables are repeated due to excision constraints. 
\begin{equation}
\label{dirk}
    \beta_k \sim Dirichlet_{|V|}(\bm{\eta} \odot \bm{b_k})
\end{equation}

where $\bm{\eta} = (\eta_1, \dots, \eta_{|V|})$ is a hyper-parameter and notation $\odot$ signifies element-wise vector multiplication. 
Equation \ref{dirk} also highlights non-identifiability issues when $|C|$ is not minimum.
For example, consider an excision graph $G=(V,E)$ where $V=\{v_1,v_2\}$ and $E=\{(v_1,v_2)\}$ and a non-parsimonious encoding that has a variable for each excision: $C=\{b_{kv_1},b_{kv_2}\}$. 
Setting $b_{kv_1}=1$ and $b_{kv_2}=1$ is equivalent to $b_{kv_1}=0$ and $b_{kv_2}=0$.
In general, consider a clique of size $L$.
Selection of any subset of $C$ such that $|C|>1$ results in the same degeneracy of $\beta_k$.


The model for the SME composition of a specific sample describes both the distribution over SMEs and a mapping between junction reads and SMEs.
The proportion of SMEs in sample $i$ is modelled by $\theta_i$, which follows a $K$ dimensional Dirichlet distribution.
The $k^{th}$ dimension represents the probability of observing an excision from SME $k$. 
\begin{equation*}
\label{eq:gentheta}
\theta_{i} \sim Dirichlet_K(\bm{\alpha})
\end{equation*}
where $\bm{\alpha}= (\alpha_1, \dots, \alpha_K)$ are hyperparameters. 

Sample $i$ has $J_i$ observations of junction reads that overlap one or more excisions. 
The assignment of junction read $j$ in sample $i$ to a SME is denoted by $z_{ij} \in \{1, \dots, K\}$ and follows a Multinomial distribution.
\begin{equation*}
\label{eq:genz}
z_{ij} \sim Multinomial(\theta_{i})
\end{equation*}
The data likelihood is represented by observed random variables $w_{ij}$, modelling the $j^{th}$ junction read in sample $i$ ($w_{ij} \in \{1, \dots, |V|\}$)  and follows a Multinomial distribution.

\begin{equation*}
\label{eq:genw}
w_{ij} \sim Multinomial (\beta_{z_{ij}})
\end{equation*}
Here, the parameter for the Multinomial is the $\beta$ selected by variable $z_{ij}$.

\subsection{Inference Algorithm}
\label{sec:inf}
We develop a Gibbs sampling algorithm to fit our model that 
 proceeds by first sampling parameter values from their priors.
Then, for each latent variable $z$, we sample from their complete conditionals, i.e., the probability of $z$ given all other random variables in the model (see \S \ref{sec:supp_gibbs}).
To determine Gibbs sampling convergence, we used Relative Fixed-Width Stopping Rules (RFWSR)~\parencite{flegal2015relative}.
RFWSR sequentially checks the width of a confidence interval relative to a threshold (here, $\sigma = 0.001$) based on the effective sample size. 
Most variables yield efficient updates (full derivations are provided in the supplemental materials \S \ref{sec:supp_gibbs}); however, special consideration is required for $z$ and $b$.


\subsubsection{Sampling \texorpdfstring{$\bm{z}$}{\textbf{z}}.}
Gibbs sampling can be inefficient, particularly for admixture models where the likelihood computation requires iterating over the full, typically high dimensional data~\parencite{hoffman2013stochastic}.
We improved the efficiency of our inference algorithms by exploiting the low dimensionality of junction reads.
In admixture modelling, 
the dimension of the data is typically much larger than the number of distinct data items in a sample. 
E.g., in topic modelling, the number of words in the vocabulary is much larger than the number of unique words in a document. 
However, in this context, the number of distinct excisions in a gene is much fewer than the size of the total number of reads.

We can exploit the fact that we treat the expression of junction reads as draws from a Multinomial.
Given probabilities $p_1, p_2, \dots, p_K$ such that $\sum_{i = 1}^{K}p_i = 1$ and $S$ as the number of draws, the naive algorithm for sampling from a discrete distribution divides the interval $[0,1]$ into $K$ segments with the length equal to $p_1, p_2, \dots, p_K$. 
A number is sampled from the Uniform distribution $\mathcal{U}(0,1)$, and the matched category is found using binary search; this procedure is repeated $S$ times.
The sampling algorithm requires $\mathcal{O}(K)$ time for initialization, and then $\mathcal{O}(S \log(K))$ time for sampling~\parencite{startek2016asymptotically}. 
In our setting, for a given gene and excision, $K$ is the number of SMEs and the number of draws, $S$, is the number of times the excision appeared in the gene. 
So, using this scheme to sample the latent variable for SME assignment yields a
 complexity of $\mathcal{O}(K + S \log(K))$, compared with $\mathcal{O}(SK + S\log(K))$ which saves significant time when $S >> K$. 

\subsubsection{Sampling \texorpdfstring{$\bm{b}$}{\textbf{b}}.}
\label{sec:algorithm}

Each iteration of Gibbs sampling requires sampling $\bm{b_k}$, which is non-trivial since the distribution of $\bm{b_k}$ is defined over independent sets of $G$. 
At each iteration we perform a local search among valid configurations (independent sets) in $\Omega$ using a novel local search algorithm for independent sets. 
At iteration \textit{t}, given the current configuration $\bm{b_k}^{t}$ and $\beta_k^{t}$ we select $\Phi = \{\phi_1, \phi_2, \dots, \phi_T\}$ valid configurations (Alg. \ref{alg:localindsearch}), among which we sample according to a Multinomial distribution (Sec. \ref{sec:supp_localsearch}).
\begin{equation*}
    \bm{b_k}^{t+1}\sim Multinomial(\phi_1^{t}, \phi_2^{t}, \dots, \phi_T^{t})
\end{equation*}
After Gibbs Sampling converges, \BREM{} collapses SMEs with the same excision configuration.

\subsubsection{Bounding the number of SMEs.}
In order to guarantee the constraints are respected, we need to compute a lower bound on the number of SMEs. 
The minimum number of SMEs is equal to the chromatic number of $G$. 
Since $G$ is an interval graph this number is the same as the number of vertices in the maximum clique ($K$). 
Therefore, we set the minimum number of SMEs to $K$ such that there exists at least one SME for each $b$ variable. 

\subsection{Differential SMEs}
Here, we define our generalized linear model (GLM) to compute differential SME usage using the fitted model parameters in \BREM{}.
We quantify differential SME usage based on the expression profile across all SMEs for $2$ groups of samples in each gene. 
The $z$ variables represent the mapping of an excision observation to a SME. 
Let the counts across all unique junction reads for sample $i$ be denoted $z_{i}$.
Then, we can express $z_{i}$ as a Dirichlet-multinomial
\begin{equation*}
    z_{i1},\dots,z_{iJ_i} \sim DirMult \left(\sum_j z_{ij}, \alpha \odot p_i \right)
\end{equation*}
where $p_{ij} = \frac{\exp(x_i \beta_j + \mu_j)}{\sum_k \exp(x_i \beta_k + \mu_k)}$. We set $\alpha \sim \gamma(1 + 10^{-4}, 10^{-4})$ to stabilize maximum likelihood estimation~\parencite{li2018annotation}.
Finally, to test differential SMEs between two groups, we construct two models: (a) a DirMult GLM where we set $x_i=0$ for one group and $x_i=1$ for the other and (b) a DirMult GLM where all $x_i=0$.
Differential SMEs are quantified by a likelihood ratio test with $K-1$ degrees of freedom, where $K$ is the number of SMEs.



\section{Results}
\label{sec:results}

All transcript annotation-free AS characterization methods must first reconstruct spliced transcripts based only on aligned RNA-seq data and approximate gene starting and ending coordinates. 
Here, we consider four state-of-the-art methods for AS characterization: rMATS~\parencite{shen2014rmats}, LeafCutter~\parencite{li2018annotation}, Cufflinks~\parencite{trapnell2010transcript}, and StringTie~\parencite{pertea2015stringtie}.
These four methods range from single splicing event to full-length transcripts and so we refer to their reconstructed output collectively as \textit{transcript segments}.
SMEs can be interpreted as the sequence of mRNA excisions within a transcript segment. 
Since these methods have different targets for reconstruction, comparing them presents a number of challenges.
First, transcript segments must be mapped to a reference annotation to evaluate reconstruction accuracy.
Computed full-length transcripts may include or exclude a subset of excisions or differ slightly in excision coordinates.
Methods that consider single splice events may produce many slightly different versions of the same excision and are attempting to solve a less complex problem than full-length transcript reconstruction.
Second, each method computes different abundance measures that may be unavailable to competing methods (e.g., FPKM is poorly defined for singular splicing events).
Therefore, we develop two measures for matching computed transcript segments of any size to a reference set of transcripts and focus on the evaluation of differential splicing for whichever transcript segment is produced by each method.

\subsection{Evaluation Criteria}
To appropriately evaluate methods that compute transcript segments of varying size we define two measures based on excision matching.
If the set of expressed transcripts is known \textit{a priori}, computed excisions can be evaluated using variants of homogeneity scores and partial precision and recall~\parencite{aguiar2018bayesian}, however that is not the case here.
Since we can compute excisions from exons, but not vice versa, we define transcript segments by the set of their component excisions.
Reconstructed transcript segments may include exons from disparate transcripts.
With a known reference, we compute the number of excisions that appear in any reference transcript and normalize by the total number of excisions. 
We define the $k^{th}$ computed transcript segment $T_k$ as a subset of excisions, or, $T_k \subseteq \{1,\dots,V\}$. 
Let the set of reference transcripts be $T^t$
and the $v^{th}$ excision be $e_v$.
The set $T^t$ either represents a simulated baseline or known experimental transcripts from a well characterized cell type. 
The \textit{partial homogeneity score (phs)} for transcript $T_k$ in sample $i$ can be computed as 
\begin{equation*}
    s_i^{phs}(T_k)=max_{T \in T^t}  \frac{\sum_{e_j \in T_k} \mathbbm{1}\left[e_j \in T \right]}{  |T_k| }
\end{equation*} 
where $\mathbbm{1}[e_j  \in T]$ is $1$ if $e_j$ matches an excision in $T$ and $0$ otherwise. 
Here, we consider two excisions $e_v$ and $e_w$ as matching if the donor and acceptor splice sites of $e_v$ are at most $6$ bases from the donor and acceptor splice sites of $e_w$. 
The score $s_i^{phs}$ enforces that excisions are sampled from the same true transcript and is normalized by the size of the computed transcript segment.
We also define $\hat{s}_i^{phs}$, which normalizes computed transcript segments by the true transcript length.
\begin{equation*}
    \hat{s}_i^{phs}(T_k)=max_{T \in T^t}  \frac{\sum_{e_j \in T_k} \mathbbm{1}\left[e_j \in T \right]}{ |T| }
\end{equation*} 
Both scores $s_i^{phs}$ and $\hat{s}_i^{phs}$ are related to the Jaccard index but importantly emphasize different goals.
Score $s_i^{phs}$ will tend to produce better scores for methods that compute shorter transcript segments; as long as the shorter transcript segments are accurate (they are contained within true transcripts), this score will be close to $1$.
In contrast, $\hat{s}_i^{phs}$ prefers longer transcript segments and will be close to $1$ if the computed transcript is both accurate and full-length.
Either $s_i^{phs}$, $\hat{s}_i^{phs}$, Jaccard index, or some linear combination thereof can be used depending on the goals of the study.

Finally, let the set of computed transcripts be $T_{(i)}^c=\{T_k\}$. 
An overall score for sample $i$ can then be computed as
\begin{gather*}
    s_i^{phs}=\frac{\sum_{T_k \in T_{(i)}^c} s_i^{phs}(T_k)}{|T_{(i)}^c|} \qquad \text{and} \qquad  \hat{s}_i^{phs}=\frac{\sum_{T_k \in T_{(i)}^c} \hat{s}_i^{phs}(T_k)}{|T_{(i)}^c|}
\end{gather*}

To compute precision and recall, we first match computed transcript segment $T_k$ to the true transcript $T \in T^t$ with maximum $s_i^{phs}$ or $\hat{s}_i^{phs}$.
Let the matched transcript be $T^*$.
Then, we label each excision $e_j \in T_k$ as a true positive (TP) if $1\left[e_j \in T^* \right]=1$ and a false positive (FP) otherwise. 
Excisions are labeled as false negatives (FN) if they exist in a true transcript but were not included in any computed transcript. 
The F1 score is computed as the harmonic mean of precision and recall, which are computed as: $precision=\frac{TP}{TP+FP}$ and $recall=\frac{TP}{TP+FN}$.




\subsection{Data}
\label{sec:sim-res}

\subsubsection{Simulations}

We evaluate isoform reconstruction with extensive simulations from the Polyester simulator~\parencite{frazee2015polyester}.
We consider protein coding genes from reference chromosomes of the GENCODE comprehensive gene annotation version V34 (human genome version GRCh38/hg38)~\parencite{frankish2019gencode}. 
We generated a diverse set of genes by randomly sampling from GENCODE until we had at least $60$ genes in each of the following categories of transcript counts (isoforms) $ \in \{2, 3, 5, 7, 10, 15, 25\}$,  ($420$ genes in total).
For each gene, we simulated $800$ samples at $50x$ coverage and then downsampled each gene to produce new datasets of $25x$ and $5x$ coverage. 
The samples were simulated using $8$ groups of $100$ samples each with different fold changes ($1$, $1$, $1$, $1.1$, $1.25$, $1.5$, $3$, and $5$) to allow for estimation of false discoveries and differential splicing sensitivity~\parencite{li2018annotation}.
The number of reads varied per sample based on a negative binomial distribution for read counts~\parencite{frazee2015polyester}.
The output FASTA files from Polyester were aligned to the human genome (version GRCh38/hg38) using STAR aligner with default parameters and GENCODE v34 annotations~\parencite{dobin2013star}.
In total, we simulated $1260 (= 420 genes \times 3 coverages)$ genes yielding over a million BAM files. 
Data simulation steps are detailed in Sec.~\ref{sec:supp_data_sim} (See Fig. \ref{fig:data_sim} for additional information).

\subsubsection{Experimental Data}

To evaluate our differential SME model, we consider both bulk and single-cell sequencing experimental data.
The GEUVADIS data contains bulk RNA-seq data from lymphoblastoid cell lines in $465$ individuals.
The samples provided from this data set are ethnically diverse and span five populations: Utah residents with northern and western European ancestry (CEU), Finnish from Finland (FIN), British from England and Scotland (GBR), Toscani from Italia (TSI), and Yoruba from Ibadan, Nigeria (YRI); each population consists of $89-95$ samples.
In our differential SME analysis, we group the CEU, FIN, GBR, and TSI populations into European (EUR) and classified the YRI population as African (AFR). 

We also consider single-cell data from the European Genome-Phenome Archive (EGA).
This data was used to investigate the response of monocytes to bacterial and viral stimuli in two populations, each with $100$ males self-reported as having predominately African ancestry (AFB) or European ancestry (EUB) within Ghent Belgium~\parencite{rotival2019defining}.
Up to five samples from peripheral blood mononuclear cells were collected for each individual resulting in $970$ total samples. 
One sample remained untreated, while the four other samples were exposed over 6 hours to bacterial lipopolysaccharide (LPS), synthetic triacylated lipopeptide (Pam$_3$CSK$_4$), imidazoquinoline compound (R848), and human seasonal influenza A virus (IAV).
We compared the untreated samples with the group of treated samples to evaluate differential SMEs.

\subsection{Preprocessing}
\label{sec:preprocessing}
The input to our model is the set of junction reads mapped to a reference genome. 
In this work, we map reads to the reference genome using STAR aligner (V. 2.7.3a). 
We used gene annotations whenever possible, including in STAR alignments since some methods require gene annotations and STAR highly recommends using them when available. 
Gene annotations were also used to generate the BAM files for the GEUVADIS data. 
However, gene annotations are not required to  execute \BREM{}.


\subsubsection{Excision Extraction} 
\label{sec:intron}

We build a model for each gene independently based on approximate gene starting and terminal coordinates.
Given approximate gene coordinates, we extract reads in each sample that overlap excisions (junction reads) using RegTools (version 0.5.1). 
The intervals of genes that overlap are combined. 
We refined the set of excisions by removing reads that do not map uniquely (e.g., due to paralogous genes), short excisions ($<50bp$), long excisions ($>500,000bp$), and false positive splice junctions identified by Portcullis~\parencite{mapleson2018efficient}. 
The extracted junction from the mapped reads form the input to \BREM{}. 



\subsection{Model Selection}

\label{sec:exp-res}

\BREM{} assumes that the number of SMEs ($K$) is given as input. 
However, the number of SMEs should not be less than the chromatic number of the excision interval graph, or, equivalently, the maximum independent set ($IS$) of the complement graph.
Using the interval graph property, we can compute the chromatic number in polynomial time. 
Then, for each gene, we trained our model with $K = IS + x$, where $x \in \{0,2,4,6,8,10,12,14,16\}$ and selected  the model with the highest \textit{predictive likelihood}. 
Predictive likelihood is commonly used to perform model selection on admixture and topic models~\parencite{wallach2009evaluation} and is less prone to overfitting than likelihood. 
To select hyperparameters, we implemented a grid search on held-out genes where $\alpha \in \{0.001, 0.01, 1, 5, 10\} $, $\eta \in \{0.01, 1, 5, 10\}$, $r$ and $s \in \{1, 5, 10\} (r = s)$. 
Throughout the subsequent experiments, we set $\eta = 0.01$ and $\alpha=r=s=1$ for both simulated and experimental data. 
For convergence, we check RFWSR every 50 iterations after burn-in (500 iterations) and stop sampling after 100 iterations if RFWSR$<\sigma$.





\subsection{Transcript Reconstruction} 
\label{trecon}
We applied \BREM{}, Cufflinks, LeafCutter, StringTie, and rMATS to reconstruct transcripts, splice events, or SMEs in all $1260 (420 \times 3)$ simulated genes.
Before computing precision, recall, and F1 score, the computed transcript segments must be matched to true transcripts; here, we quantify this using the partial homogeneity scores~(Fig.~\ref{fig:phsf}).

\begin{figure}[h]
    \centering
    \includegraphics[trim={0 0 0 0}, clip, width=0.85\textwidth]{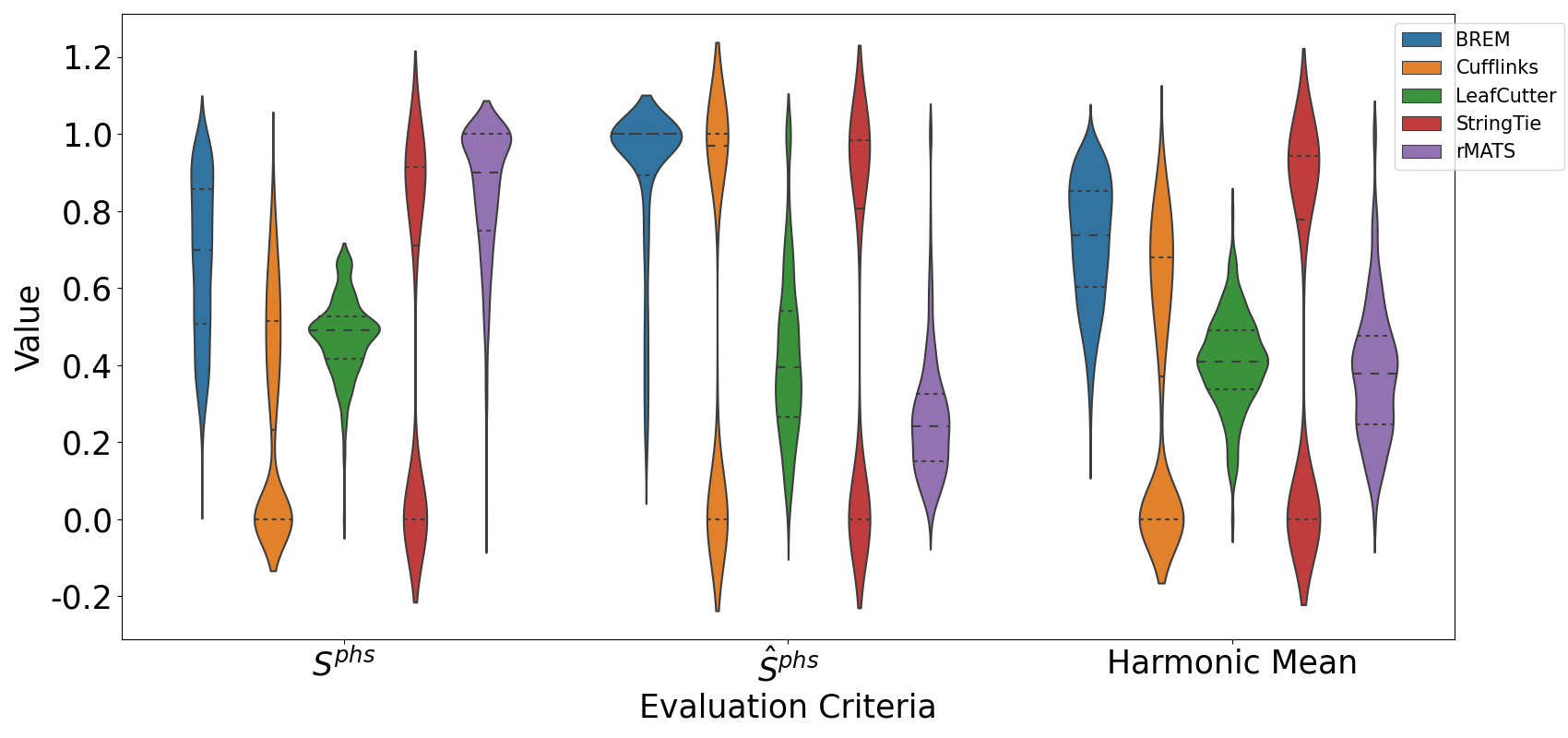}
    \caption{\textbf{Transcript segment matching to reference.} Violin plots for $s^{phs}$, $\hat{s}^{phs}$ and their harmonic mean across five methods in the simulated data. The horizontal lines show the quartiles in each of the plots.}
    \label{fig:phsf}
\end{figure}


\noindent LeafCutter and rMATS match true transcripts well when normalizing by the number of excisions in the computed transcript segments ($s^{phs}$); however, when normalizing by the true transcript, the $\hat{s}^{phs}$ score for both methods predictably decreases due to the size of transcript segments produced.
Since Cufflinks and StringTie both aim to reconstruct full-length transcripts, they perform comparatively well when normalizing by the size of the true transcript; however, Cufflinks score dramatically decreases when normalizing by the size of the computed transcript, indicating that its computed transcript lengths in terms of excisions, are inaccurate.

\begin{figure}
    \centering
    \includegraphics[trim={0 0 0 0}, clip, width=0.75\textwidth]{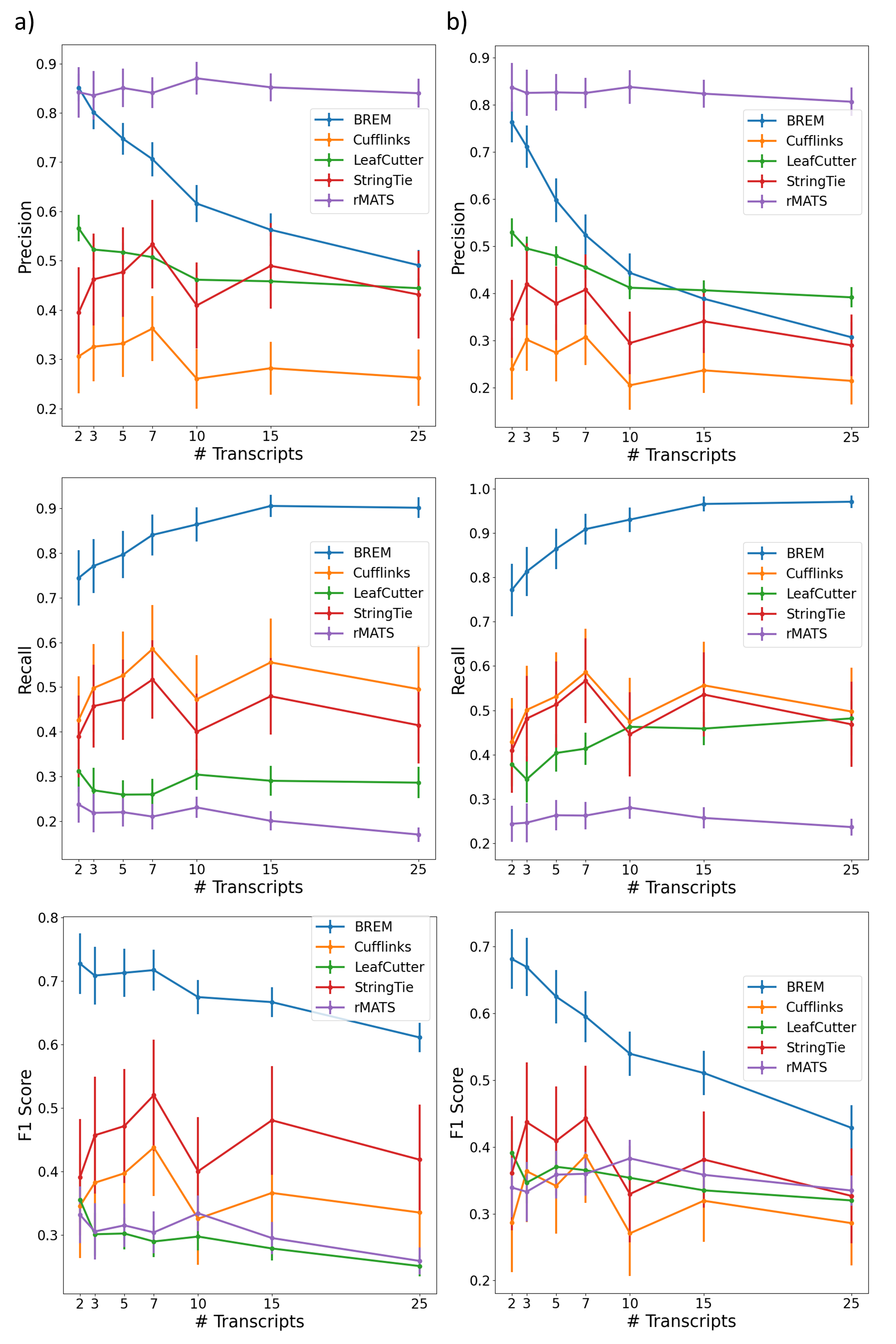}
    \caption{\textbf{Performance on simulated data.} Precision, recall and F1 Score on simulated data for \BREM{} (blue), Cufflinks (orange), LeafCutter (green), StringTie (red) and rMATS (purple) based on \textbf{(a)} $S^{phs}$ and \textbf{(b)} $\hat{S}^{phs}$.}
    \label{fig:prf1}
\end{figure}

Interestingly, StringTie does not suffer from the same significant decrease as Cufflinks, though both StringTie and Cufflinks exhibited high variance.
In comparison, \BREM{} demonstrated far less variability in $s^{phs}$ and $\hat{s}^{phs}$ than Cufflinks and StringTie, while maintaining high performance. 

Next, to evaluate the impact of the parameter that controls the number of SMEs ($K$), we varied $K$ from the chromatic number in the excision interval graph (equivalently, size of the maximum independent set ($IS$) in the complement graph) to $IS + 16$.
The trend for both $s^{phs}$~(Fig.~\ref{fig:fig1}, top) and $\hat{s}^{phs}$~(Fig.~\ref{fig:fig1}, bottom) are similar: as $K$ increases, precision increases initially and then remains flat while recall decreases monotonically. 
This is likely due to two factors.
First, \BREM{} collapses SMEs with the same excision configuration after convergence. 
This means that the \textit{effective K} is much lower when $K$ is much larger than the number of alternative transcripts. 
Second, \BREM{} benefits from the flexibility of additional SMEs initially, but eventually when $K>>IS$, \BREM{} learns SMEs that are low abundance and noisy.

Having matched computed transcripts with true transcripts, we next evaluated each method with respect to precision, recall, and F1 score for the top match using $s^{phs}$~(Fig.~\ref{fig:prf1}a) and $\hat{s}^{phs}$~(Fig.~\ref{fig:prf1}b).
First, rMATS is highly selective, exhibiting high precision regardless of the length of the transcript.
This, however, is to be expected since rMATS scores consistently high when considering $s^{phs}$, but also consistently low when considering $\hat{s}^{phs}$.
Since rMATS is concerned only with singular splicing events, in either case, the task is less difficult.
On the other hand, LeafCutter performs some local assembly of splicing events into clusters and thus has a more difficult assembly task than rMATS, though performs similarly in terms of F1 score.
Both Cufflinks and StringTie exhibit high variance, but perform considerably better than the local splicing methods in terms of recall. 
\BREM{}, situated between these extremes, achieves higher precision for most genes than the full-length transcript methods and substantially higher recall and F1 with lower standard errors.
We also tested precision, recall, and F1 score as a function of the complexity of the overlap graph (defined by the number of edges).
For genes yielding complex graphs ($|E| > 200$), \BREM{} achieves the highest recall and F1 Score, while rMATS is the most precise~(Fig.~\ref{fig:prf_nf}). 
Importantly, this shows that \BREM{} performs well when there is substantial overlap among the transcripts.
As a function of the number of excisions, \BREM{} also achieves the highest recall (Fig \ref{fig:prf_nodes}). 
The flexibility of our admixture modelling allows \BREM{} to focus on producing high confidence transcript segments rather than fixing the size to be small (e.g., individual splice events) or large (full-length transcript).



Next, we tested the sensitivity of \BREM{} to model parameters; in particular, we tested how the mean posterior SME length (denoted $|SME|$ and defined by the number of excisions in a SME) varied as a function of $K$, $r$, and $s$. We trained the model setting $r, s \in \{0.1, 1, 10, 100\}$ and $K \in \{IS, IS + 5, IS + 10, IS + 15\}$.
The parameter $K$ did not correlate with $|SME|$, likely due to \BREM{} collapsing posterior SMEs with the same excision usage. 
However, as we increased the prior mean of $Beta(r,s)$, $|SME|$ also increased~(Fig.~\ref{fig:sensitivity_rs}).
This is consistent with the interpretation of $r$ and $s$ in the model: $r$ and $s$ control the prior probability of including an excision in SMEs.
As the mean of $Beta(r,s)$ increases, larger SMEs become more likely in the posterior.
However, this relationship is not strictly monotonic, as other model parameters, properties of the transcripts, and stochasticity of model inference interact with the effect of $r$ and $s$ on $|SME|$. 
\BREM{} is also fast, with the running time increasing linearly as a function of $K$, $|C|$, and the average number of junction reads across samples (Fig. \ref{fig:runtime_2}). 


\subsection{Differential Expression in Simulated Data}
Each simulated gene consisted of $8$ groups of $100$ samples with fold changes $1$, $1$, $1$, $1.1$, $1.25$, $1.5$, $3$, and $5$.
We computed differential expression for each method and all pairwise groupings of the samples ($28$ in total).
Pairwise comparisons between the first three groups enabled estimation of false discoveries. 
We randomized the processing order of genes and allocated each method a full week on a $128$ core computer to processes the simulated data (Table~\ref{tab:ds}).
StringTie, rMATS, LeafCutter, and \BREM{} all finished in less than a day, while Cufflinks only finished $21.4\%$ of configurations.
Additionally, recent comparisons have shown higher ability to detect differential splicing for rMATS, StringTie, and LeafCutter when compared to Cufflinks~\parencite{li2018annotation,shen2014rmats,pertea2015stringtie}; thus, we excluded Cufflinks from the comparison.
\BREM{} achieved the highest sensitivity and accuracy of identifying differential usage (of SMEs), though StringTie achieved relatively high sensitivity with an impressively high specificity ($0.996$).

\subsection{Differential Expression in Experimental Data} 
We applied \BREM{} and our differential SME model to both GEUVADIS and EGA datasets.
After filtering genes expressed at low levels and those without conflicts in the excision interval graph, we applied \BREM{} to infer SMEs in $3983$ and $4278$ genes in the GEUVADIS and EGA data respectively.
Using our results on the precision and recall for \BREM{} with varying $K$~(Fig.~\ref{fig:fig1}), we set $K=IS+4$.
We then applied our Dirichlet Multinomial model to compute differential SME usage across the two data sets. 
We used the super population (African vs. European) to group samples in GEUVADIS and treatment status in the EGA dataset.
After multiple comparisons correction using Benjamini-Hochberg~\parencite{benjamini1995controlling}, p-values were well-calibrated~(Fig.~\ref{fig:exp}) and we observed $2105$ and $1961$ genes with significant differential SME usage in the GEUVADIS and EGA data (FDR corrected $p<0.05$). 

\begin{figure}
    \centering
    \includegraphics[trim={20 30 0 10},width=0.60\textwidth]{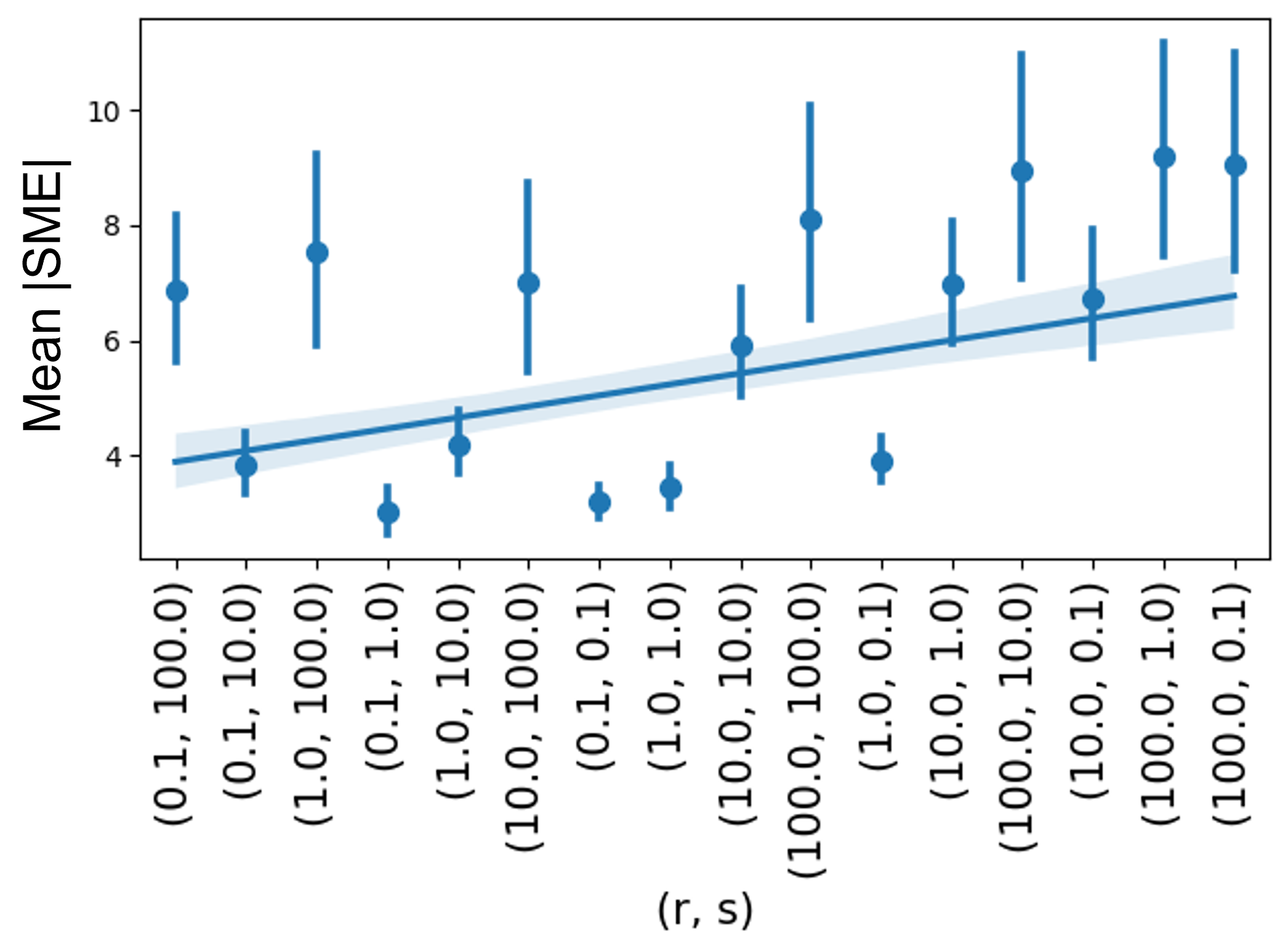}
    \caption{\textbf{Sensitivity analysis of SME size with respect to the parameters $r$ and $s$.} X-axis depicts combinations of $\pi$ variable prior parameters, ordered by $\pi$ mean, i.e., $\frac{r}{r+s}$. In y-axis, we compute the average SME size.}
    \label{fig:sensitivity_rs}
\end{figure}

\begin{table}[h!]
    \centering
    \begin{tabular}{|c|c|c|c|}
    \hline
      \textbf{Method}  & \textbf{Accuracy} & \textbf{Sensitivity}  & \textbf{Specificity}  \\ \hline
      StringTie  & $0.253$ & $0.164$ & $\bm{0.996}$ \\ \hline
        rMATS & $0.130$ & $0.0284$ & $0.990$ \\ \hline
        LeafCutter & $0.131$  & $0.0292$ & $0.980$ \\ \hline
        \BREM{} & $\bm{0.303}$  & $\bm{0.233}$ & $0.889$ \\ \hline
    \end{tabular}
    \caption{\textbf{Differential Splicing Results on Simulated Data.}}
    \label{tab:ds}
\end{table}

We conducted a gene ontology (GO) analysis using genes with differential SME expression as the target list and all genes input into \BREM{} as the background list~\parencite{eden2009gorilla}.
In the GEUVADIS data, the top $12$ GO terms in the Biological Process ontology ranked by p-value (p $<2.97 \times 10^{-6}$) referenced regulation of biomolecular processes.
This is consistent with a growing body of evidence that suggests splicing plays a major role in regulating gene expression~\parencite{gehring2020anything,gutierrez2015tissue} and metabolism~\parencite{kozlovski2017role,annalora2017alternative,qiao2019comprehensive}.
In the Molecular Function ontology, alternative splicing plays an integral role in the top $18$ GO terms, which reference ATP, DNA, drug, and other molecular binding (p $<5.69 \times 10^{-5}$)~\parencite{sciarrillo2020role,ji2020silico}.
In the EGA data, both molecular function and biomolecular processes exhibited significant associations with regulation of and binding to kinase proteins (GO:0046330, GO:0043507, GO:0046328, GO:0019901; p $<9.7 \times 10^{-4}$).
Alternative splicing is known to (a) regulate the binding of kinase proteins~\parencite{kelemen2013function} and (b) increase kinase protein  diversity~\parencite{anamika2009functional}.

\subsection{Novel Splice Junctions, SMEs and Transcripts}
We quantified the total number and percentages of novel versus known splice junctions and SMEs or transcripts in both GEUVADIS and EGA datasets. 
Since our processing pipeline focuses on excisions, and is thus similar to LeafCutter, and we are testing reconstruction, we compared our results to only Cufflinks and StringTie. 
We only consider SMEs that are expressed in $10$ or more samples, where the $k^{th}$ SME is considered expressed in sample $i$ if there exists $10$ or more $z_{ij}=k$. 
We allowed inferred junction locations to differ by at most $6$ nucleotide bases from the reference to be considered matching. 
We followed the recommended pipelines for StringTie and Cufflinks and merged per-sample assemblies.
We considered an SME or computed transcript as novel if it was not a subset of an annotated transcript (and known otherwise). 
Splice junctions are considered novel if they do not exist in the reference (and known otherwise). 

All methods produce far fewer novel SMEs, transcripts, and splice junctions in GEUVADIS compared to the EGA data (Fig. \ref{fig:novel}, a and b).
This may be due to the reference transcripts of lymphoblastoid cell lines being more well characterized than monocytes or due to the differences in sequencing platforms (bulk versus single cell RNA-seq).
\BREM{} and Cufflinks produced larger proportions of known to novel splice junctions and SMEs or transcripts in GEUVADIS whereas StringTie produces far more novel transcripts (Fig. \ref{fig:novel}, c and d).  
This discrepancy was larger in the EGA data, where StringTie produced much higher proportions of novel transcripts ($<5\%$ were observed in the reference).


\section{Discussion}
\label{sec:con}


In this work, we presented \BREM{}, a new hierarchical \textbf{B}ayesian admixture model for the \textbf{r}econstruction of \textbf{e}xcised \textbf{m}RNA and quantifying differential SME usage.
The structure of our graphical model depends on both modelling assumptions of the data generating process and an interval graph that encodes relationships between excisions (for which we prove is parsimonious).
We develop efficient inference algorithms based on a polynomial time computation of node covers and local search over independent sets.
We presented the new problem formulation of SME reconstruction, which interpolates between the local splicing and full-transcript views of alternative splicing.
To enable the comparison between local splicing, full-transcript, and excision based methods, we developed two partial homogeneity scores to match computed transcript segments to reference annotations. Finally, we demonstrated that \BREM{} is accurate in terms of SME reconstruction and identifying differential expression when compared to four state-of-the-art methods on simulated data and that it captures relevant biological signal in bulk and single-cell RNA-seq data. 

There are several interesting directions for future work, both in terms of SME modelling and addressing limitations of \BREM{}.
First, a natural extension to the parametric admixture model presented here, comes from placing nonparametric priors on both the individual specific ($\theta$) and global ($\beta$) SME distributions~\parencite{teh2006hierarchical}.
An immediate benefit to this Bayesian nonparametric modelling is that model selection of $K$ is integrated with the inference algorithm; also, the complexity of the model can adapt with new samples, for example, from a different tissue or disease condition.
Second, explicit modelling of the count-based nature of single cell RNA-seq data could, in theory, be accommodated with varying data likelihoods in our probabilistic model.
Third, although we defined our differential SME model based on expression profiles across SMEs in a gene, it could similarly be constructed to test differential SMEs within a gene. 
Fourth, SME-QTLs are a natural analog to splicing QTLs (sQTLs)~\parencite{gtex2015genotype} and transcript ratio QTLs (trQTLs)~\parencite{lappalainen2013transcriptome}, but may require extensive experimental validation to evaluate.
Lastly, reads covering a single exon could be incorporated to  improve abundances estimation, model allele specific expression, or detect alternative transcription start or end sites and retained introns (all of which cannot be detected by \BREM{} due to its focus on excised mRNA).  

\section*{Data Availability}

Data from the GEUVADIS project is available through ArrayExpress database (www.ebi.ac.uk/arrayexpress) under accession number E-GEUV-6. 
Data from the EGA project is available through European Genome-Phenome Archive (ega-archive.org) under Study ID: EGAS00001001895 and Dataset ID: EGAD00001002714.
The source code for \BREM{} is freely available at \href{https://github.com/bayesomicslab/BREM}{https://github.com/bayesomicslab/BREM}.

\section*{Competing interest statement}
The authors declare no competing interests.

\section*{Acknowledgements}
We are gracious to the GEUVADIS and EGAS00001001895 projects for providing open and easy access to experimental data.
MH, DM, and DA were funded by DA's University of Connecticut start-up research funds.

\printbibliography

 \end{refsection}
\newpage

 \begin{refsection}
\onecolumn

\section*{Supplementary Materials}
\label{sec:supp}
\addcontentsline{toc}{section}{Appendices}
\renewcommand{\thesubsection}{S\arabic{subsection}}
\renewcommand{\thefigure}{S\arabic{figure}}
\setcounter{figure}{0} 
\setcounter{subsection}{0} 

\subsection{Related Work}

\label{sec:rw}
Methods for AS characterization can be grouped into three categories based on their assumed input annotations: no reference annotations (\textit{de novo} assembly), gene transcripts and their exon composition (transcript annotation-based), and gene starting and ending positions only (transcript annotation-free).
\textit{De novo} transcriptome assembly methods, like Trinity~\parencite{haas2013novo} and ABySS~\parencite{birol2009novo}, compute transcripts from unaligned sequence reads, typically without the benefit of reference annotations. 
When a reference genome sequence is well characterized, transcript annotation-based and annotation-free methods have been shown to produce more accurate transcripts and quantifications~\parencite{marchant2016comparing}; since our focus is on species with well-categorized genome sequences, we restrict our attention to methods that assume sequences reads can be mapped to a genome reference.

Transcript annotation-based and annotation-free isoform reconstruction methods begin by aligning RNA-seq reads to a reference genome using a splice-aware aligner~\parencite{langmead2012fast,dobin2013star}. 
The overwhelming majority of these methods reconstruct full-length transcripts as ordered sets of exons, focusing on the RNA that is retained.
The Bayesian isoform discovery and individual specific quantification (BIISQ) method models transcript reconstruction with a nonparametric Bayesian hierarchical model, where samples are mixtures of transcripts sampled from a population transcript distribution~\parencite{aguiar2018bayesian}.
While BIISQ was shown to have high accuracy on low abundance isoforms, it requires both the genes and the composite exon coordinates, and is unable to construct isoform transcripts that deviate from this reference annotation.
Cufflinks and StringTie are two methods that construct full-length transcripts and can operate both with or without transcript annotations. 
Cufflinks reconstructs transcripts as minimum paths in an associated graph, where the aligned reads are vertices, and edges denote the compatibility of isoforms~\parencite{trapnell2010transcript}.
StringTie models transcript reconstruction using maximum network flow on a splice graph, where paths and read coverage inform isoform composition and quantification respectively~\parencite{pertea2015stringtie}.
Both are well-established state-of-the-art methods, but consider samples individually during the initial reconstruction.
Additionally, all aforementioned methods are restricted to constructing full-length isoforms, a problem that is made challenging by exon boundaries that are difficult to identify and variability in read depths across transcripts.


A more recent class of isoform reconstruction and quantification methods focus on characterizing local splicing events.
The local splicing and hierarchical model rMATS detects differential usage of exons through the comparison of exon-inclusions in junction reads among five different alternative splicing events~\parencite{shen2014rmats}. 
Interestingly, LeafCutter focuses on the mRNA that is excised rather than the constituent exons of a transcript to identify local splicing events. 
First, LeafCutter computes local splicing events from RNA-Seq data then constructs a graph $G_L = (V_L,E_L)$ where vertices, $V_L$, are excisions and edges, $E_L$, connect excisions that share a donor or acceptor splice site~\parencite{li2018annotation}.
Subsequently, differential splicing of excised sequences in the connected components of $G_L$ is computed using a Dirichlet-Multinomial generalized linear model on read counts.
LeafCutter does not suffer the same disadvantages of methods that use exonic sequences or attempt to reconstruct full-length transcripts, though at the expense of the inability to identify certain splicing events like alternative transcription start sites. 
These methods also may fail to capture interactions between splicing events on the same transcript and may conflate transcripts that share splice events. 
For example, if two transcripts share an excision, the read counts on the shared junction will be summed conflating the two transcripts and potentially masking differential expression across two populations (Figure~\ref{fig:overview} C).

Our method, \BREM{}, is situated in between full-length transcript and local splicing methods~(Fig.~\ref{fig:overview}).
It benefits from the transcript annotation-free nature of excisions, while also being able to support both local splicing events, full-length transcripts, and variable lengths in between.
\BREM{} also shares similarities with BIISQ in that it considers all samples jointly and is defined as a formal probabilistic model, enabling the quantification of uncertainty and direct interpretation of fitted model parameters that are used to both explore the results and develop a method for differential testing. 


\subsection{Additional Model Details}
\label{supmoddets}
\subsubsection{Notations}

Variables and indices, parameters and hyper-parameters and sets are as following:
\begin{itemize}
    \item $V$ is the set of unique excisions, indexed by $v$ and its size of denoted by $|V|$.
    \item $N$ is the number of samples and are indexed by $i$.
    \item $J_i$ 
    is the number of excisions in $i$th sample. 
    The excisions of a sample are indexed by $j$. But the length of all the samples are not necessarily the same. Furthermore, some samples might not have some of the excisions from the set of unique excisions $V$. 
    \item $K$ is the number of sequences of mRNA excisions (SMEs). 
    For the $j$th excision in the $i$th sample ($i\in \{1, \dots,N\}$ and $j \in \{1, \dots, J_i\}$), we assign an SME $k \in \{1, \dots, K\}$. 
    \item Graph $G = (V, E)$, where $V$ is the set of excisions and there is an edge between two excisions \textit{iff} their intersection is non-empty.
    \item $\Omega$ is the set of all the independent sets in $G$.
    \item $\mathcal{N}_v = \{u|\{u, v\} \in E(G)\}$ is the set of all the neighbors of node $v$ in the interval graph $G$.
    \item $\phi^{it}_{k}$ is the selected configuration as SME $k$ in the iteration $it$ and follows a Multinomial distribution \\ ($ \sim Multinomial (\phi_{k1}, \phi_{k2}, \dots, \phi_{kt}, \dots, \phi_{kT})$).
    \item $C$ is the set of all Bernoulli random variables required for encoding all conflicts in $G(V,E)$, and $|C|$ is equal to minimum node cover in $G$.  
    \item Hyper-parameter $\bm{\alpha} = (\alpha_1, \dots, \alpha_K)$ is a $K$-dimensional vector and prior for $\theta$ variable.
    \item For the $i$th sample, variable $\theta_i \sim Dirichlet_K(\bm{\alpha})$ is a $K$-dimensional Dirichlet distribution and represents the proportions of the SMEs in sample $i$. 
    So $\bm{\theta}$ is a $N \times K$ matrix such that each row shows the distribution of SMEs for a sample and $\theta_{ik}$ shows the proportion of SME $k$ in sample $i$ ($\bm{\theta} \in \mathbb{R}^{N \times K}$).
    \item Variable $z_{ij}$ is the SME assignment for $j$th excision in $i$th sample. It can take a natural value between $1$ and $K$ and follows a Multinomial distribution ($\bm{Z} \in \{1, \dots, K\}^{N \times J}$ and $z_{ij} \sim Multinomial(\theta_i)$).
    \item Hyper-parameters $r$ and $s$ are priors for $\pi$ Beta distribution. 
    \item Variable $\pi_k \sim Beta(r,s), \forall k=\{1, \dots, K\}$, so $\bm{\pi}$ is a $K$-dimensional vector and prior for Bernoulli variable $\bm{b}$.
    \item Hyper-parameter $\bm{\eta} = (\eta_1, \dots, \eta_{|V|})$ is a $|V|$-dimensional vector and prior for $\beta$ variable.
    \item For SME $k$, $\beta_k \sim Dirichlet_{|V|}(\bm{\eta} \odot \bm{b_{k}})$ is a $|V|$-dimensional Dirichlet which represents the distribution of the SME $k$ over the excisions. $|V|$-dimensional vector $\bm{b_{k}} = (b_{k1}, \dots, b_{k|V|})$ (also written as $\bm{b_{k.}}$) is the $k$th row of the $\bm{b}$ matrix and collects the Bernoulli variables for all the unique excisions. 
    The \textit{dot} in $\bm{b_{k.}}$ means all the unique excisions in row $k$. 
    Notation $\odot$ is element-wise multiplication. 
    Matrix $\bm{\beta}$ is $K \times |V|$ and the element in $k$th row and $v$th columns shows the proportion of excisions $v$ in SME $k$, so matrix $\bm{\beta} \in \mathbb{R}^{K \times |V|}$. 
    Note that the Bernoulli variables can turn off/on certain dimensions of $\beta$ variables. 
    \item In the $i$th sample, the $j$th excision is $w_{ij}$ and is observed and follows a Multinomial distribution ($w_{ij} \sim Multinomial(\beta_{z_{ij}})$), so matrix $\bm{W}$ is a $N \times J$ ($W \in \{1, \dots, |V|\}^{N \times J}$) and $w_{ij}$ is the element in $i$th row and $j$th column of $\bm{W}$ and is $j$th excision in $i$th sample and is observed. 
    Note that in $\bm{W}$, row $i$ correspond to sample $i$, but not all the rows have the same number of columns due to the differences between the number of excisions in different samples, \emph{i.e.} row $i$ has exactly $J_i$ columns (elements) which correspond to the excisions in the sample $i$. We call $\bm{W}$ here as a matrix for the ease of notation, but it is actually a list of list.
    The same explanation applies to matrix $\bm{Z}$ too.
    \item $\oplus$ is exclusive OR.
    \item $\odot$ is element-wise vector multiplication.
\end{itemize}
\newpage

\subsubsection{Graphical Model}
\begin{figure}[h]
\centering
\includegraphics[width=0.95\textwidth]{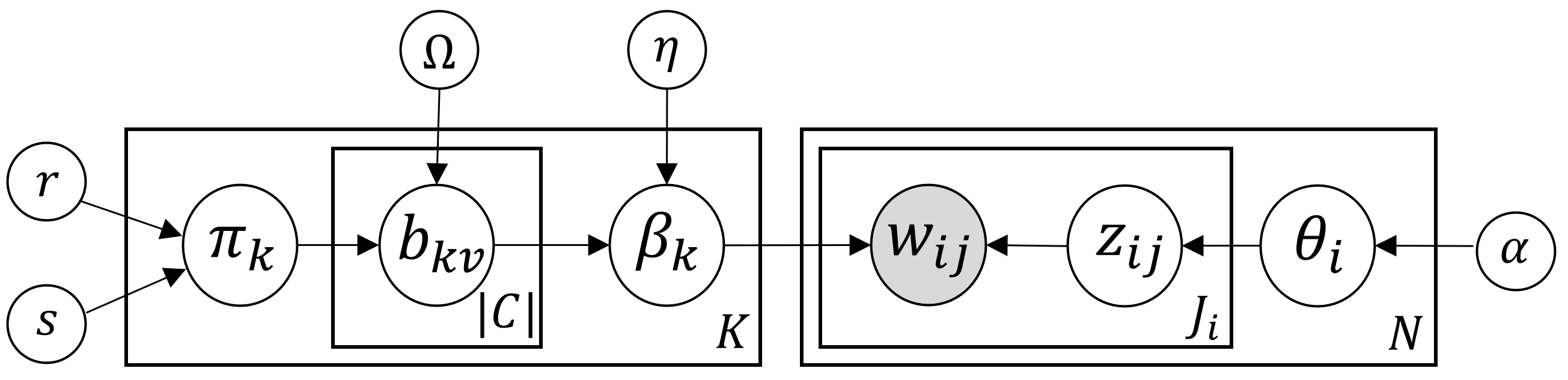}
\caption{Graphical model for \BREM{}. The variables $\bm{\pi}$, $\bm{b}$, and $\bm{\beta}$ control the global sequence of mRNA excisions (SME) structure, while $\bm{w}$, $\bm{z}$, and $\bm{\theta}$ control the sample-specific distribution of SMEs.}
\label{fig:graphicalmodel}
\end{figure}

\begin{align*}
   \theta_i  & \sim Dirichlet_K(\bm{\alpha}), &\forall i \in \{1, \dots, N\}\\
   z_{ij}  & \sim Multinomial(\theta_i), & \forall i \in \{1, \dots, N\}, \forall j \in \{1, \dots J_i\}\\
  w_{ij}  & \sim Multinomial(\beta_{z_{ij}}), &\forall i \in \{1, \dots, N\}, \forall j \in \{1, \dots J_i\}\\
  \beta_k & \sim Dirichlet_{|V|}(\bm{\eta} \odot \bm{b_{k}}), &\forall k \in \{1, \dots, K\}\\
    b_{kv} & \sim Bernoulli(\pi_k), &\forall k \in \{1, \dots, K\}, \forall v \in |C|\\
    \pi_k & \sim Beta(r,s), &\forall k \in \{1, \dots, K\}
\end{align*}

In the calculations, B(.) in Dirichlet distribution is
\begin{align*}
   B(\bm{\alpha}) &= \frac{\prod_{i=k}^K \Gamma(\alpha_i)}{\Gamma(\sum_{i=1}^K \alpha_i)}
\end{align*}

in which $\bm{\alpha}$ is the vector of concentration parameters and $K \ge 2$ is the number of SMEs 
in Dirichlet. Gamma function $\Gamma(n)=(n-1)!$.

$\bm{C}$ is a $N \times K$ matrix, in which $c_{ik}$ is the number of excisions in the $i$th sample that have been assigned to SME $k$. 

$\bm{\lambda}$ is a $K\times |V|$ matrix, in which $\lambda_{kv}$ is the number of times excision $v$ has been assigned to SME $k$.

\subsection{Inference - Gibbs Sampling}
\label{sec:supp_gibbs}
We fist compute the complete conditionals of all the variables in the model and then sample the variables according to order described in this section.

\subsubsection{Complete Conditional of \texorpdfstring{$\bm{\theta_i}$}{Ti}}

\begin{align}
   p(\theta_i|\bm{\alpha}, z_{i, j=1:J_i})  & \propto p(z_{i, j=1:J_i}, \bm{\alpha}, \theta_i) \nonumber \\
            &= \prod_{j=1}^{J_i} p(z_{ij}|\theta_i) p(\theta_i|\bm{\alpha}) \label{eq:theta2}
\end{align}
The first term in Eq. \ref{eq:theta2} is a Multinomial and the second term is a $K$-dimensional Dirichlet. So for updating $\theta_i$ we have:
\begin{align*}
   p(\theta_i|\bm{\alpha}, z_{i, j=1:J_i})  & \propto \frac{B(\bm{\alpha}+\bm{c_{i.}})}{B(\bm{\alpha})} \times Dirichlet_K (\bm{\alpha}+\bm{c_{i.}})
\end{align*}

where $\bm{c_{i.}}$ is a $K$-dimensional vector containing the proportion of SMEs in sample $i$. 

\begin{align*}
\bm{c_{i.}} = [c_{i,k=1}, c_{i, k=2}, \dots, c_{i, k=K}]
\end{align*}

\subsubsection{Complete Conditional of \texorpdfstring{$\bm{Z}$}{\textbf{Z}}}
Lets consider complete conditional of one variable $z_{ij}$ (SME assignment of $j$th excision in $i$th sample):

\begin{align*}
   p(z_{ij}|\theta_i, w_{ij}, \bm{\beta_{1:K}})  & \propto p(z_{ij}, w_{ij}, \theta_i, \bm{\beta_{1:K}})\\
   & \propto p(w_{ij}|z_{ij},\bm{\beta_{1:K}}) p(z_{ij}|\theta_i)
\end{align*}
The notation $\bm{\beta_{1:K}}$ means that variable $z_{ij}$ is dependent on $\beta_1$ to $\beta_K$. 
Since $z_{ij} \in \{1, 2, \dots, K\}$ (discrete random variable), complete conditional for an assignment $z_{ij} = k$ would be

\begin{align*}
   p(z_{ij} = k|\theta_i, w_{ij}, \bm{\beta_{1:K}})   &= \frac{p(z_{ij}=k| \theta_i) p(w_{ij}| z_{ij}=k, \bm{\beta_{1:K}})}{\sum_{k=1}^K p(z_{ij}=k| \theta_i) p(w_{ij}| z_{ij}=k, \bm{\beta_{1:K}})}
\end{align*}

And in sample $i$, for SME $k$:

\begin{align*}
   p(z_{ij} = k|\theta_i, w_{ij}, \bm{\beta_{1:K}}) & \propto p(z_{ij}=k| \theta_i) p(w_{ij}| z_{ij}=k, \bm{\beta_{1:K}})\\
   & \propto \theta^{c_{ik}}\theta^{\alpha_i-1} \times \prod_{v=1}^{|V|} \beta_{kv}^{\lambda_{kv}} \beta^{\eta_{kv}b_{kv}-1}
\end{align*}


The probability of assigning the (unique) excision $v$ (in any position $j$ in sample $i$) to SME $k$:

\begin{align*}
   p(z_{ij} = k|\theta_i, w_{ij} = v, \bm{\beta_{1:K}})   &= \frac{\theta_{ik} \beta_{kv}}{\sum_{k=1}^{K}\theta_{ik} \beta_{kv}}
\end{align*}

\subsubsection{Complete Conditional of \texorpdfstring{$\bm{\beta}$}{\textbf{B}}}
For SME $k$, $\beta_k$ is a $|V|$-dimensional Dirichlet distribution over the unique excisions. We used Bernoulli variables $b$ that restrict the number of excisions in one SME (the size of $\beta_k$ variables) by defining $\beta_k \sim Dirichlet_{|V|}(\eta_1 b_{k1}, \dots, \eta_{|V|}b_{k|V|})$. So $\bm{\beta}$ is a degenerate Dirichlet distribution.

\begin{align*}
  p(\beta_k|\bm{W}, \bm{Z}, \bm{b_{k.}}) & \propto p(\beta_k, w_{..}, z_{..}, \bm{b_{k.}})\\
  &= p(w_{..}|z_{..},\beta_k) p(\beta_k| \bm{b_{k.}},\bm{\eta})
\end{align*}

For SME $k$:
\begin{align*}
  p(\beta_k|\bm{W}, \bm{Z}, \bm{b_{k.}}) & \propto \prod_{v=1}^V \beta_{kv}^{\lambda_{kv}} \times \frac{\Gamma (\sum_{v=1}^V \eta_v b_{kv}}{\prod_{v=1}^V \Gamma(\eta_v b_{kv})} \times \prod_{v=1}^V \beta_{kv}^{\eta_v b_{kv}-1} \\
  & \propto \prod_{v=1}^V \beta_{kv}^{\lambda_{kv}} \times \beta_{kv}^{\eta_v b_{kv}-1} \\
  & \propto Dir_{|V|}(\bm{\lambda_{k.}}+\bm{\eta} \odot \bm{b_{k.}})
\end{align*}
In which vector $\bm{\lambda_{k.}}$ is the count of excisions that have been assigned to SME $k$:

\begin{equation*}
    \bm{\lambda_{k.}} = [\lambda_{k, v=1}, \dots, \lambda_{k, v=|V|}]
\end{equation*}



\subsubsection{Complete Conditional of \texorpdfstring{$\bm{b}$}{\textbf{b}}}




Let the constrained space be $\Omega$, which spans over the set of all independent sets of excisions' interval graph $G$ (See Section \ref{sec:modeldescription}). 
Here, we consider computing the Gibbs updates for independent sets $\{\Phi_1,\dots,\Phi_T\} \in \Omega$.
For ease of exposition, we consider $b_{kv}$ given a configuration $\hat{\Phi}$.
For an excision $v$, we compute the probability of occurrence of $v$ in SME $k$. This probability is obtained by the complete conditional of $b_{kv}$. 
Note: For computing $b_{kv}$, we need to consider relevant dimensions of Dirichlet.
For example, in a SME $k$, for the calculation of complete conditional for $b_{kv} = 1$, such dimensions include all the excisions that are not in the neighborhood of $v$ ($\{v'|v'\notin \mathcal{N}_v\}$).
$\mathcal{N}_v$ is the set of all the neighbors of $v$, and does not include $v$ itself (open neighborhood).
We denote $\bm{b^{(-kv)}}$ as the vector of $b$ variables with $b_{kv}$ removed and suppress hyperparameters for readability when appropriate.
\begin{align*}
  p(b_{kv}=1| \bm{\beta}, \bm{\pi}, \bm{b^{(-kv)}},\bm{W}, \bm{Z}, \bm{\theta}) \propto p(\bm{\beta}, \bm{\pi}, \bm{b},\bm{W}, \bm{Z}, \bm{\theta})\\
  \propto p(b_{kv}=1| \beta_k,\pi_k,\bm{b_{k.}})\\
    = p(b_{kv}=1|\pi_k) p(\pi_k|r,s)p(\beta_k|\bm{b_{k.}}, \bm{\eta}) \label{eq:gibbsb1}\\
   = p(b_{kv} = 1|\pi_k) \frac{\Gamma(r+s)}{\Gamma(r)\Gamma(s)}  \pi_k^{r-1} (1 - \pi_k)^{s-1} \\ \frac{\Gamma(\sum_{i \in \hat{\Phi} \cup \{b_{kv}\}} \eta_i b_{ki})}{\prod_{i \in \hat{\Phi} \cup \{b_{kv}\}}\Gamma(\eta_i b_{ki})}  \prod_{i \in \hat{\Phi} \cup \{b_{kv}\}} \beta_{ki}^{\eta_i b_{ki} - 1}\\
     \propto p(b_{kv}=1|\pi_k)  \pi_k^{r-1} (1 - \pi_k)^{s-1}  \\ \frac{\Gamma(\sum_{i \in \hat{\Phi} \cup \{b_{kv}\}} \eta_i b_{ki})}{\prod_{i \in \hat{\Phi} \cup \{b_{kv}\}}\Gamma(\eta_i b_{ki})}  \prod_{i \in \hat{\Phi} \cup \{b_{kv}\}} \beta_{ki}^{\eta_i b_{ki} - 1}\\
     \propto \pi_k  \pi_k^{r-1} (1 - \pi_k)^{s-1} \\ \frac{\Gamma(\sum_{i \in \hat{\Phi} \cup \{b_{kv}\}} \eta_i b_{ki})}{\prod_{i \in \hat{\Phi} \cup \{b_{kv}\}}\Gamma(\eta_i b_{ki})}  \prod_{i \in \hat{\Phi} \cup \{b_{kv}\}} \beta_{ki}^{\eta_i b_{ki} - 1}\\
        \propto  \pi_k^{r} (1 - \pi_k)^{s-1}   \frac{\Gamma(\sum_{i \in \hat{\Phi} \cup \{b_{kv}\}} \eta_i b_{ki})}{\prod_{i \in \hat{\Phi} \cup \{b_{kv}\}}\Gamma(\eta_i b_{ki})}  \prod_{i \in \hat{\Phi} \cup \{b_{kv}\}} \beta_{ki}^{\eta_i b_{ki} - 1}
 \end{align*}
which is the product of a $Beta(r+1,s)$ and a degenerate Dirichlet.
In SME $k$, for computing complete conditional for $b_{kv} = 0$ we need to involve the other excisions except $v$, so:

\begin{align*}
  p(b_{kv}=0| \bm{\beta}, \bm{\pi}, \bm{b^{(-kv)}},\bm{W}, \bm{Z}, \bm{\theta})  \propto p(\bm{\beta}, \bm{\pi}, \bm{b},\bm{W}, \bm{Z}, \bm{\theta}) \\
  \propto p(b_{kv}=0| \beta_k,\pi_k,\bm{b_{k.}}) 
   \propto  p(b_{kv}=0|\pi_k)p(\pi_k|r,s)p(\beta_k|b_{k.}, \eta) \\
   = p(b_{kv}=0|\pi_k)  \frac{\Gamma(r+s)}{\Gamma(r)\Gamma(s)}  \pi_k^{r-1} (1 - \pi_k)^{s-1}  \frac{\Gamma(\sum_{i \in \hat{\Phi} \setminus \{b_{kv}\}} \eta_i b_{ki})}{\prod_{i \in \hat{\Phi} \setminus \{b_{kv}\}}\Gamma(\eta_i b_{ki})}  \prod_{i \in \hat{\Phi} \setminus \{b_{kv}\}} \beta_{ki}^{\eta_i b_{ki} - 1}\\
     \propto p(b_{kv}=0|\pi_k)  \pi_k^{r-1} (1 - \pi_k)^{s-1}  \frac{\Gamma(\sum_{i \in \hat{\Phi} \setminus \{b_{kv}\}} \eta_i b_{ki})}{\prod_{i \in \hat{\Phi} \setminus \{b_{kv}\}}\Gamma(\eta_i b_{ki})}  \prod_{i \in \hat{\Phi} \setminus \{b_{kv}\}} \beta_{ki}^{\eta_i b_{ki} - 1}\\
     \propto (1-\pi_k)  \pi_k^{r-1} (1 - \pi_k)^{s-1}   \frac{\Gamma(\sum_{i \in \hat{\Phi} \setminus \{b_{kv}\}} \eta_i b_{ki})}{\prod_{i \in \hat{\Phi} \setminus \{b_{kv}\}}\Gamma(\eta_i b_{ki})}  \prod_{i \in \hat{\Phi} \setminus \{b_{kv}\}} \beta_{ki}^{\eta_i b_{ki} - 1}\\
     \propto  \pi_k^{r-1} (1 - \pi_k)^{s}   \frac{\Gamma(\sum_{i \in \hat{\Phi} \setminus \{b_{kv}\}} \eta_i b_{ki})}{\prod_{i \in \hat{\Phi} \setminus \{b_{kv}\}}\Gamma(\eta_i b_{ki})}  \prod_{i \in \hat{\Phi} \setminus \{b_{kv}\}} \beta_{ki}^{\eta_i b_{ki} - 1}
 \end{align*}
 
which is the product of a $Beta(r,s+1)$ and a degenerate Dirichlet.
In general, we have $T$ independent sets $\{\Phi_1,\dots,\Phi_T\}$ and the update is computed by sampling 
\begin{equation}
Cat\left( \frac{p(\Phi_1)}{\sum p(\Phi_{i=1}^T)},\dots,\frac{p(\Phi_T)}{\sum p(\Phi_{i=1}^T)} \right)
\end{equation}
We develop two algorithms for computing $\{\Phi_1,\dots,\Phi_T\}$.
First, we compute $p(b_{kv}=1|\cdot)$ and $p(b_{kv}=0|\cdot)$.
Then, for every node that is a neighbor to $b_{kv}$, we know this node is currently off.
So, we compute $p(b_{ki}=1,b_{kv}=0)$ for each $i \in \mathcal{N}_v$ if setting $b_{ki}=1$ yields a valid configuration.
Second, we update SME structure by moving from one independent set to another that is 'close'.

In the interval graph of excisions $G = (V,E)$, let $\Omega$ be the set of all the independent sets and $\phi_{kt} \subseteq \Omega$ be the $t$\textsuperscript{th} locally generated valid configuration for SME $k$. We define the neighbor of $\phi_{kt}$ as  $\mathcal{N}(\phi_{kt})$, i.e. the set of the nodes that intersect with some of the nodes in $\phi_{kt}$ (or $\mathcal{N}(\phi_{kt}) = \{u| \{u,v\} \in E\text{ for some }v \in \phi_{kt}\}$). Then $p(\phi_{kt})$ is computed as the following:

\begin{align*}
  p(\phi_{kt}| \bm{\beta}, \bm{\pi}, \bm{b},\bm{W}, \bm{Z}, \bm{\theta}) \propto \pi_k^{r+|\phi_{kt}|-1} (1-\pi_k)^{s+|\mathcal{N}(\phi_{kt})|-1}  \frac{\Gamma(\sum_{i \in V\setminus \mathcal{N}(\phi_{kt})} \eta_i b_{ki})}{\prod_{i \in V\setminus \mathcal{N}(\phi_{kt})}\Gamma(\eta_i b_{ki})}  \prod_{i \in V\setminus \mathcal{N}(\phi_{kt})} \beta_{ki}^{\eta_i b_{ki} - 1} \\
 \end{align*}



\subsubsection{Complete Conditional of \texorpdfstring{$\pi_k$}{\textbf{Pi}}}

We define $m_k$ equal to the number of excisions that are selected in SME $k$, \emph{i.e.} the excisions whose corresponding Bernoulli variable is $1$ in the current Gibbs iteration:

\begin{equation*}
    m_k = \sum_{v \in V} \bm{1}[b_{kv}=1], \hspace{20pt} \forall k \in K
\end{equation*}

\begin{align*}
   p(\pi_k|\bm{b_{k.}},r,s)& \propto p(\pi_k, \bm{b_{k.}},r,s) \\
   & =p(\pi_k|r,s)p(b_{k.}|\pi_k)\\
   & =\frac{\Gamma(r+s)}{\Gamma(r)\Gamma(s)} \times \pi_k^{r-1}(1-\pi_k)^{s-1} \times \pi_k^{m_k} (1-\pi_k)^{|V|-m_k} \\
   & \propto \pi_k^{r+m_k-1}(1-\pi_k)^{s+|V|-m_k-1} \\
   &=\frac{\Gamma(r+m_k)\Gamma(s+|V|-m_k)}{\Gamma(r+s+|V|)} \times Beta(r+m_k, s+|V|-m_k)\\
   & \propto Beta(r+m_k, s+|V|-m_k)
\end{align*}




\subsubsection{Likelihood}
Likelihood defines how likely, the data is generated according to the generative model.





\begin{align*}
    p(\bm{W}|\bm{\beta}, \bm{Z}) & \propto p(\bm{W}, \bm{\beta}, \bm{Z}) \\
    & \propto p(\bm{W}|\bm{\beta}, \bm{Z})p(\bm{Z}|\bm{\theta}) \\
    & \propto \prod_{i=1}^N  \prod_{k=1}^{K} \prod_{v=1}^{|V|}  \beta_{kv}^{\xi^{(i)}_{kv}}
\end{align*}





where $\xi_{kv}^{(i)}$ is the number of times excision $v$ is assigned to SME $k$ in the sample $i$.

\subsubsection{Gibbs sampling algorithm}
The order of sampling is as follows:
\begin{align*}
   & p(\theta_i|\bm{\alpha}, z_{i, j=1:J_i})   \propto \prod_{j=1}^{J_i} p(z_{ij}|\theta_i) p(\theta_i|\bm{\alpha}) \\
   & p(z_{ij}|\theta_i, w_{ij}, \bm{\beta_{1:K}})  \propto p(w_{ij}|z_{ij},\bm{\beta_{1:K}}) p(z_{ij}|\theta_i) \\
   & p(\beta_k|\bm{W}, \bm{Z}, \bm{b_{k.}})  \propto p(w_{..}|z_{..},\beta_k) p(\beta_k| \bm{b_{k.}},\bm{\eta}) \\
   & p(b_{kv}=1| \bm{\beta}, \bm{\pi}, \bm{b^{(-kv)}},\bm{W}, \bm{Z}, \bm{\theta})    \propto p(b_{kv}=1| \beta_k,\pi_k,\bm{b_{k.}}) \\
   & p(\pi_k|\bm{b_{k.}},r,s)  \propto p(\pi_k|r,s)p(\bm{b_{k.}}|\pi_k)
\end{align*}

\subsubsection{Local search algorithm}
\label{sec:supp_localsearch}
We are given $\beta_k$, the proportion of excisions in SME $k$, $T$, the number of local independent sets, $S$, the set of nodes in SME $k$ (current configuration), $G = (V,E)$, the  interval graph of the excisions, and $\omega(\bar{G})$,  the size of maximum clique in the complement graph of $G$.
Then, Algorithm \ref{alg:localindsearch} outputs set $\Phi$, which includes $T$ local independent sets.
Since $G$ is an interval graph, independent sets can be computed efficiently~\cite{andrade2012fast}. 
The algorithm first decides whether to add or remove elements from the current configuration by sampling from a Bernoulli which is proportional to the size of current configuration. 
Then, excisions are selectively added or removed with probability proportional to $\beta_k$.
After Gibbs Sampling converges, \BREM{} collapses SMEs with the same excision configuration.
\newpage
\begin{algorithm}
\caption{Local Independent Set Search}
\label{alg:localindsearch}
        \textbf{Input:} $ \beta_k$, $T$, $S$, $G=(V,E)$, $\omega(\bar{G})$\\
        \textbf{Output:} $\Phi$
\hrule
\begin{algorithmic}[1]
\State $\Phi \gets \emptyset$
\While {$|\Phi| < T$}
    \State $r \gets \mathcal{B}ern(1-\frac{|S|}{\omega(\bar{G})})$ \Comment{Sample proportional to $\omega(\bar{G})$}
    \If{$r = 1$} 
        \State $\mathcal{N}_S \gets \{u\in V(G)| \{u,v\} \in E(G)\text{ for some }v \in V(G)\}$  \Comment{Set S neighborhood}
        \State $free \gets V \setminus (\mathcal{N}_S \cup S)$ 
        \If{$free \neq \emptyset$}
            \State $sel \gets Cat(\beta_{k,i  \in free})$ \Comment{\parbox[t]{.45\linewidth}{Among the nodes that if added, keeps S independent set, select based on their $\beta$ distribution}}
            \State $S \gets S \cup \{sel\}$ 
            \State $\Phi.append(S)$
        \EndIf
    \Else 
        \State $del \gets Cat(1-\beta_{k,i\in S})$ \Comment{\parbox[t]{.5\linewidth}{Among S, Select based on their $\beta$ distribution}}
        \State $S \gets S \setminus \{del\}$
        \State $\Phi.append(S)$
    \EndIf
\EndWhile
\end{algorithmic}
\end{algorithm}

\subsection{Additional Notes for the Minimum Node Cover Algorithm} 
\label{sec:node}
To identify the minimum node cover of an interval graph $G = (V,E)$ we followed an incremental algorithm proposed by~\cite{marathe1992generalized}. 
The proof of correctness specifies that nodes are included in the vertex cover set only if their presence is absolutely essential. 
Leveraging the properties of the interval graph, the algorithm first orders the vertices according to a PEO (Perfect Elimination Ordering) \cite{golumbic2004algorithmic} known as IG ordering \cite{ramalingam1988unified} in linear time and space on the order of $|E|$.
Then at each iteration, the minimum edge index which is connected to vertices is obtained. 
This index captures the nesting property of the maximal clique in the graph. 
Finally, by updating a weight counter associated to each nodes according to the mentioned index, the necessity of adding vertices to the minimum node cover is assessed.
Since calculating the minimum index for vertices takes linear time and for each vertex, the number of weight updates is equal to the degree of that vertex 
(overall $\mathcal{O}(\sum_{v\in V} d_v)$ where $d_v$ is the degree of vertex $v$)
, the whole time complexity of the algorithm is on the order of $\mathcal{O}(|E|)$, where $|E|$ is the cardinality of edge set in the interval graph.











\subsection{Preprocessing of Samples}
\label{sec:supp_data_sim}
In this section, we present the commands we ran in the preprocessing step.
\subsubsection{STAR}
\begin{verbatim}
STAR --runThreadN 20 --genomeDir ../genome_data/genome_index/ \
     --outFileNamePrefix ./person_${i}_ \
     --twopassMode  Basic --outSAMstrandField intronMotif \ 
     --outSAMtype BAM SortedByCoordinate \
     --readFilesIn ${1}/person_${i}_1.fa ${1}/person_${i}_2.fa
\end{verbatim}

GEUVADIS BAM files were downloaded from ArrayExpress (accession E-GEUV-6), which were generated by aligning fastq files using TopHat version 2.0.9 and human genome assembly version hg19. 
We generated the EGA and simulated data BAM files using the STAR aligner (version 2.7.3a).
In this example $\${1}$ is the directory where the files are located.
The input file 'person\_*\_1.fa' is a collection of genes for the $i^{th}$ sample on forward sequence and the second is the collection of genes on the backward sequence.
This allows us to use the twopassMode and we also used the intronMotif in order to obtain spliced alignments (XS).
Once the files have been aligned they are then separated out into the individual bam files of just one gene to work on at a time.

\subsubsection{Regtools}

\begin{verbatim}
regtools junctions extract -s 0 -a 6 -m 50 -M 500000 %s -o %s.junc  
\end{verbatim}

Regtools was used for an efficient filtering of the junctions.
Here the two \%s are the bam file and output name respectively.
On all data used in this project, EGA, Geuvadis, and simulations, we used the following flags:

\begin{itemize}
    \item -s: finds XS/unstranded flags
    \item -a: minimum anchor length into exon ($6$ bp)
    \item -m: minimum intron size ($50$ bp)
    \item -M: Maximum intron size ($500000$ bp)
\end{itemize}

\subsubsection{Portcullis}
The first step of portcullis is preparing the FASTA file of the reference genome; we present here an example used in the data simulations.

\begin{verbatim}
portcullis prep -t 20 -v --force -o %s_portcullis/1-prep/ \ 
      GRCh38.primary_assembly.genome.fa %s/%s.bam
\end{verbatim}

Portcullis was run on our simulations and both experimental results.
Here \%s is the name of the output folder and BAM file.

\begin{verbatim}
portcullis junc -t 20 -v -o %s_portcullis/2-junc/portcullis_all \
      --intron_gff %s_portcullis/1-prep/    
\end{verbatim}

The next step is to extract junctions in a GFF format. 
Here \%s refers to the name of the folder to search.

\begin{verbatim}
portcullis filt -t 20 -v -n --max_length 500000 \ 
      --min_cov 30 -o %s_portcullis/3-filt/portcullis_filtered \
      --intron_gff portcullis_all.junctions.tab
\end{verbatim}

      
Finally, we filter excisions; we only keep excisions that have a max length of $500000$ and have a minimum coverage of $30$. 
Here again \%s is points to the gene folder to search into for the input files.
After portcullis is complete, we keep excisions with a 90\% overlap between regtools and portcullis.

\subsection{Additional Data Details}

\begin{figure}[h]
\centering
\includegraphics[width=1\textwidth]{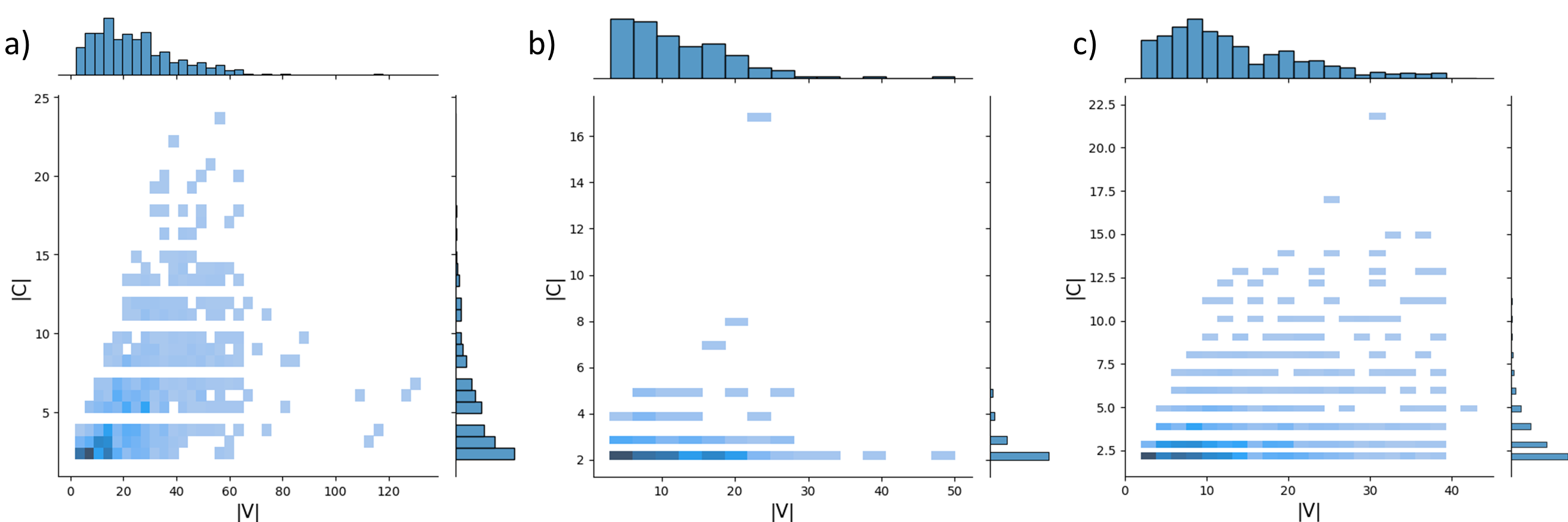}
\caption{Histogram of the minimum node cover set size with respect to the number of unique intron excisions in a) simulated data b) GEUVADIS c) EGA.}
\label{fig:data_sim}
\end{figure}

\newpage

\subsection{Additional Results on Simulated Data}

\subsubsection{\BREM{} performance as a function of \texorpdfstring{$\bm{k}$}{\textbf{k}}.}
We evaluated the impact of the parameter that controls the number of SMEs ($K$) on the precision and recall; we varied $K$ from the chromatic number in the excision interval graph (equivalently, the size of the maximum independent set ($IS$) in the complement graph) to $IS + 16$.
Fig.~\ref{fig:fig1} shows precision and recall for $s^{phs}$ and $\hat{s}^{phs}$ in top and bottom.

\begin{figure}[h]
    \centering
    \includegraphics[trim={0 0 0 0}, clip, width=0.75\textwidth]{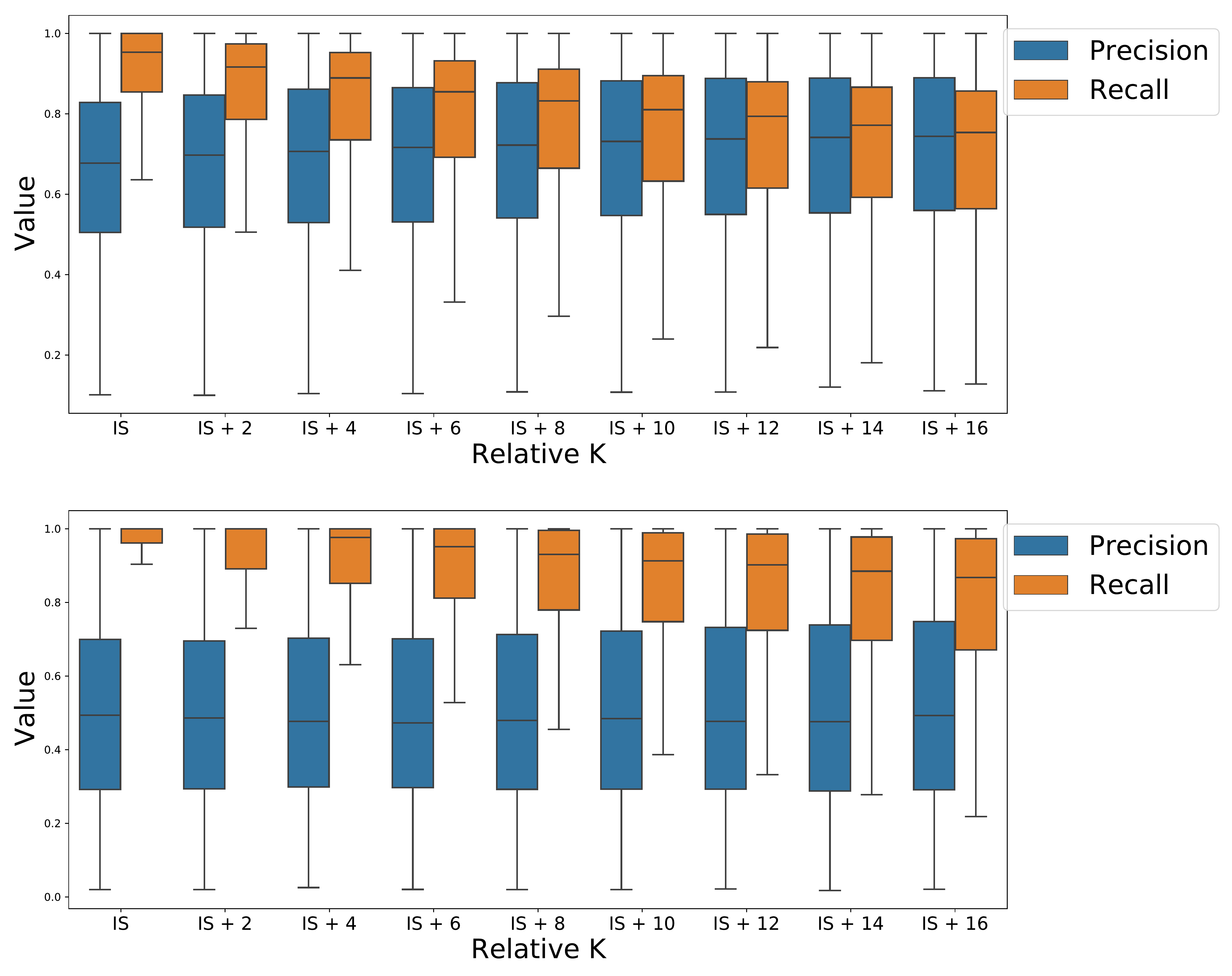}
    \caption{Precision and recall for $s^{phs}$ (top) and $\hat{s}^{phs}$ (bottom) in models where $k = IS + x$, $x \in \{2, 4, 6, 8, 10, 12, 14, 16\}$ and $IS$ is the size of maximum independent set in the complement  of the interval graph.}
    \label{fig:fig1}
\end{figure}

\subsubsection{Performance of the model when alternative transcripts having substantial overlap in sequence content}

We evaluated the performance of the methods when there are substantial overlap in the interval graph. 
We considered genes yielding complex graphs, i.e., the number of excision overlaps exceed 200 ($|E| > 200$). 
Fig.~\ref{fig:prf_nf} shows the performance metrics on for different methods.

\begin{figure}[h]
\centering
\includegraphics[width=0.95\textwidth]{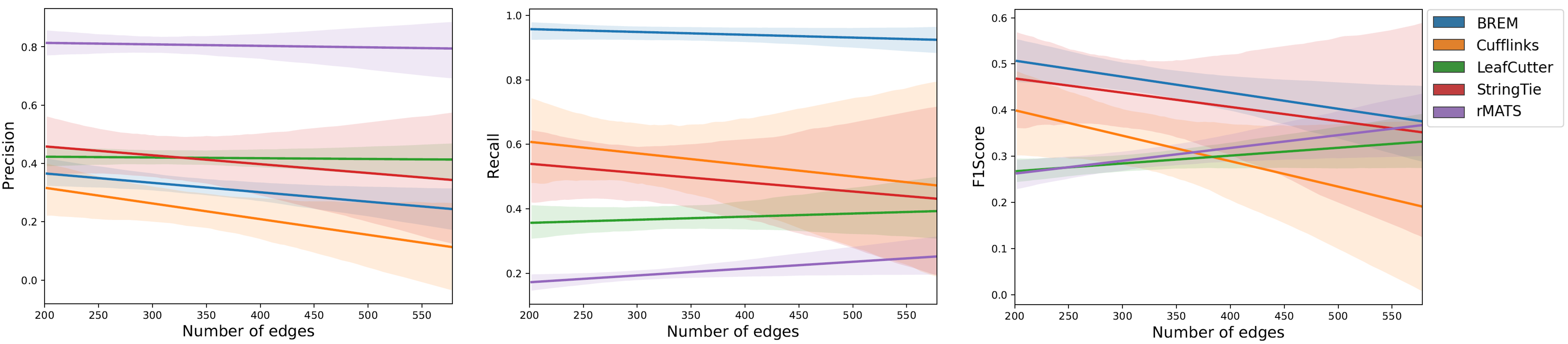}
\caption{Performance in complex genes where the number of edges exceed $200$. The x-axis shows the number of edges in the interval graph and the y-axis is the performance metric. }
\label{fig:prf_nf}
\end{figure}

\newpage

\subsubsection{Performance of the methods as function of number of unique intron excisions.}

\begin{figure}[h]
\centering
\includegraphics[width=0.95\textwidth]{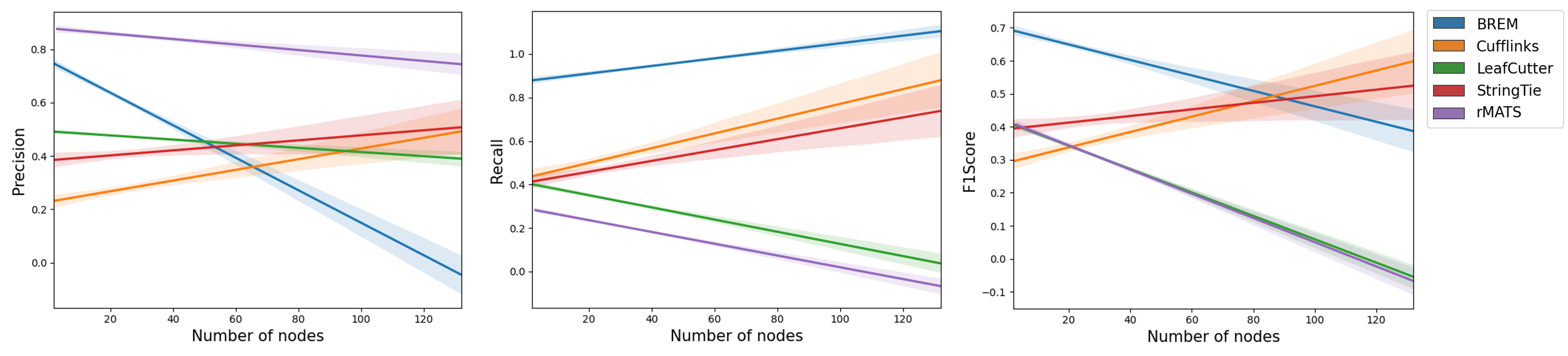}
\caption{Performance as a function of number of excisions.The x-axis shows the number of nodes in the interval graph (unique intron excisions) and the y-axis is the performance metric. }
\label{fig:prf_nodes}
\end{figure}

\subsubsection{\BREM{} running time}
Since the number of iterations varied from different runs of the same gene, we computed the running time of \BREM{} per iteration for the simulated groups of genes.
We omitted the group of $17$ genes that had an average number of junction reads that was larger than $60,000$ since this group was small relative to the sample size ($1260$ genes). 
Then we plotted the running time as a result of the number of $K$, the size of minimum node cover and the average number of junction reads (Fig~\ref{fig:runtime_2}). The model was trained on a server with Intel(R) Xeon(R) Gold 6242 @ 2.80GHz CPUs. 


\begin{figure}[h!]
    \centering
    \includegraphics[width=1\textwidth]{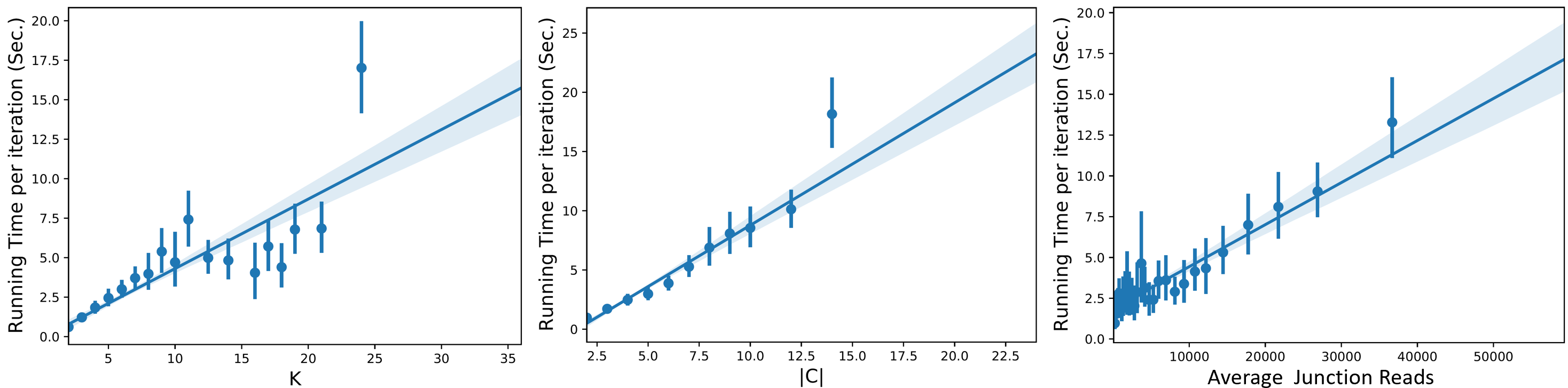}
    \caption{\BREM{} running time (Sec. per iteration) as a function of a) number of SMEs ($K$), b) size of the minimum node cover set, c) average sample size.}
    \label{fig:runtime_2}
\end{figure}

\newpage

\subsection{Additional Results on Experimental Data}

\subsubsection{P-value calibration.}
\begin{figure}[h!]
    \centering
    \includegraphics[width=1\textwidth]{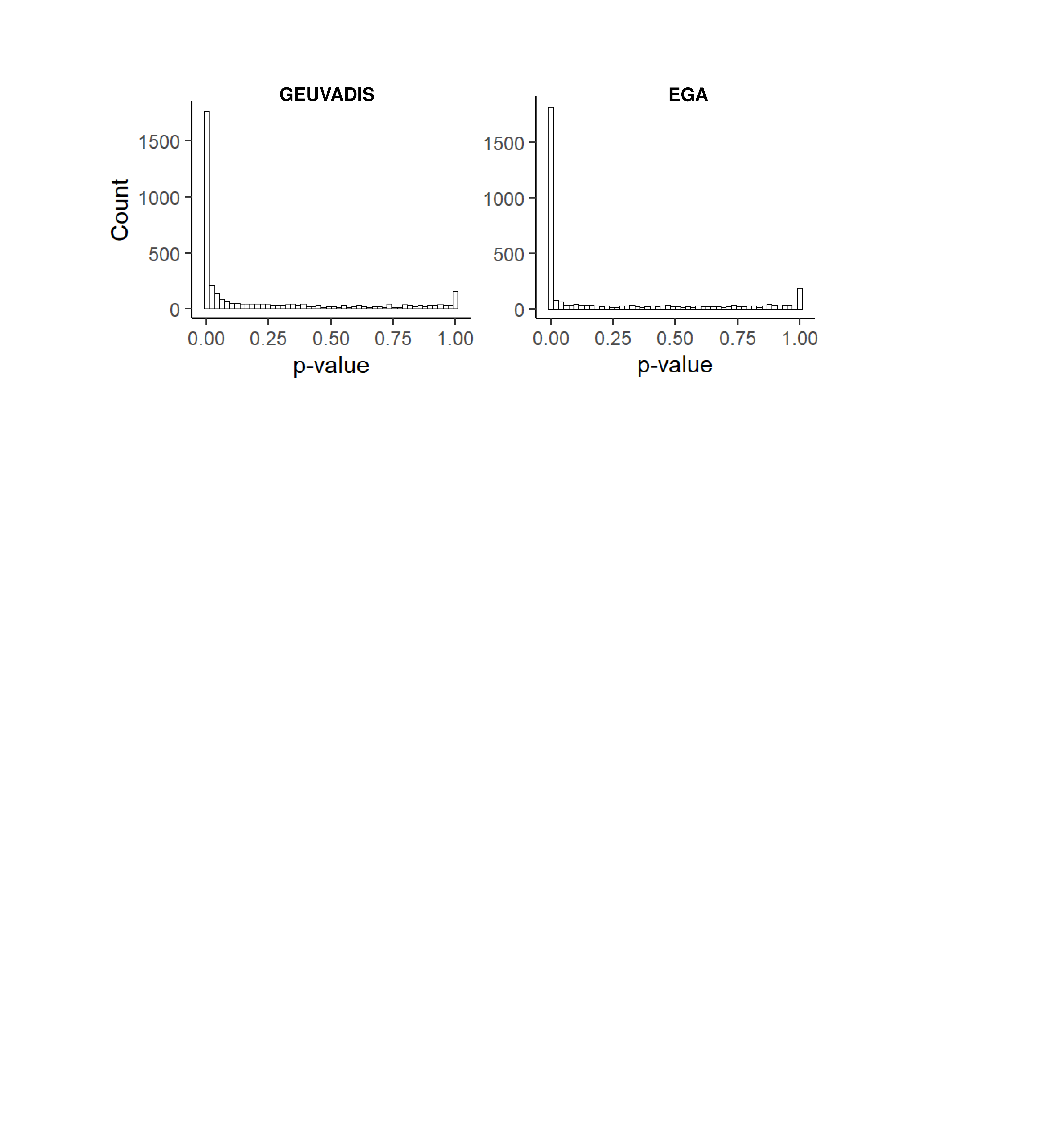}
    \caption{Differential SMEs are significantly enriched in the GEUVADIS and EGA data.}
    \label{fig:exp}
\end{figure}

\newpage
\subsubsection{Novel Introns and SMEs in Experimental data}
We computed the count and percentage of introns and SMEs that are present in or absent from the annotation reference file. 
Since our processing pipeline focuses on excisions, and is thus similar to LeafCutter, and we are testing reconstruction, we compared our results to only Cufflinks and StringTie. 
After running the methods on experimental genes, we merged single individual transcript reconstructions to produce a single file per gene for Cufflinks and StringTie.  
In BREM, a SME is expressed if it contains more than 10 junctions mapped to it. Additionally, the number of samples that express that SME is larger than 10.
Then in the merged file per gene, if a SME in \BREM{} or a transcript in StringTie and Cufflinks is a subset of any of the annotated transcripts, we count it as present, otherwise it is counted as absent. (Fig. \ref{fig:novel}, a and b)
For percentage plots, we take the percentage of present or absent for splice junctions and SMEs or transcripts separately (Fig. \ref{fig:novel}, c and d). 

\begin{figure}[h]
    \centering
    \includegraphics[trim={0 0 0 0}, clip, width=1\textwidth]{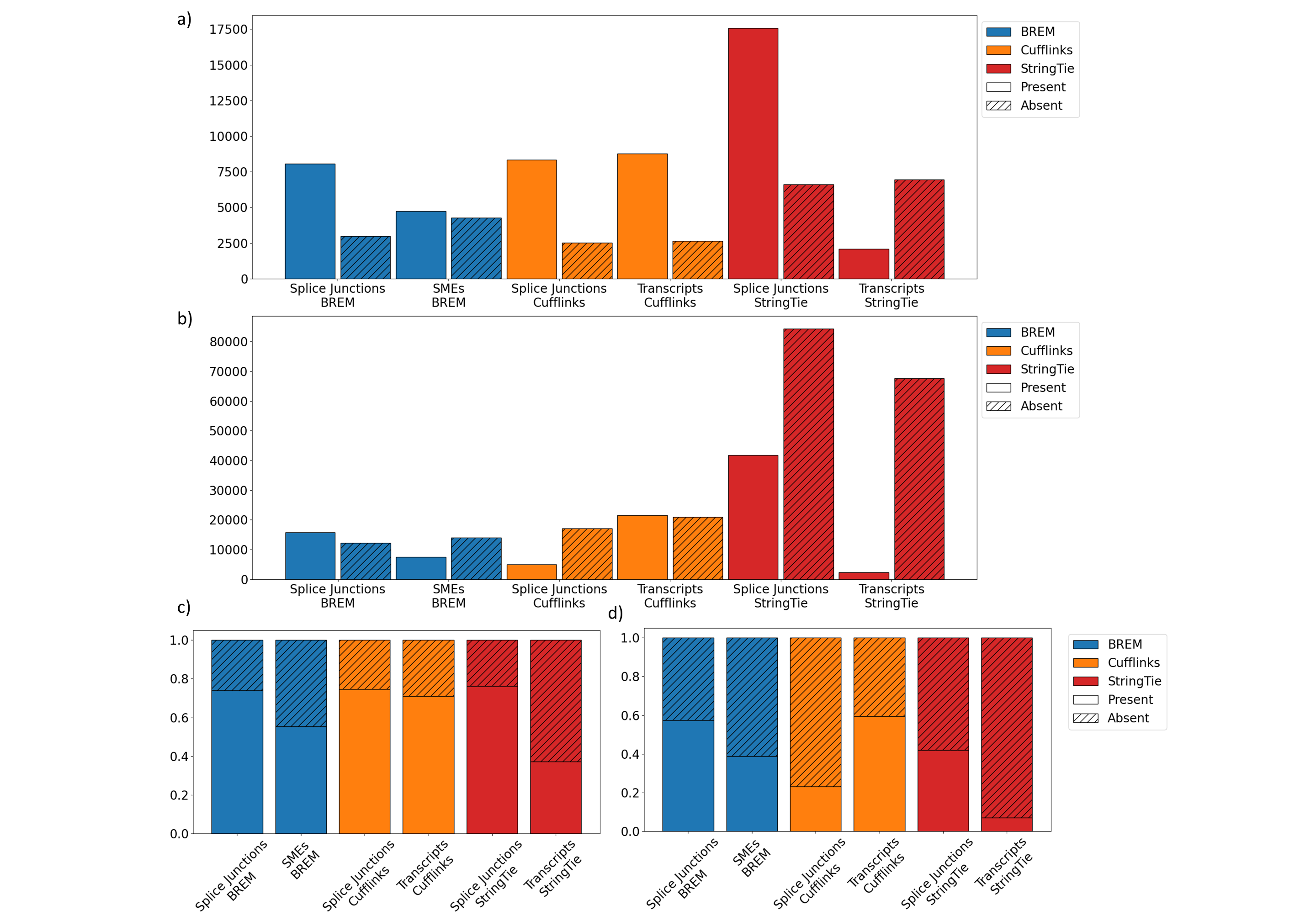}
    \caption{The count and percentage of mRNA excisions and SMEs or transcripts that are present or absent in the annotation file across \BREM{}, Cufflinks, and StringTie for experimental data. The two top plots show the counts of present and absent junctions for SMEs or transcripts in a) GEUVADIS and b) EGA. 
    The two plots in the bottom row show the percentage of present or absent junctions for SMEs or transcripts in c) GEUVADIS and d) EGA.}
    \label{fig:novel}
\end{figure}

\newpage





\printbibliography
 \end{refsection}

\end{document}